\documentclass[pra,twocolumn,amsmath,amssymb,superscriptaddress]{revtex4}

\usepackage{epsfig}
\usepackage{color}
\usepackage{graphicx}
\usepackage{bm}
\usepackage{epstopdf}

\def\beq{\begin{equation}}
\def\eeq{\end{equation}}

\newcommand{\bra}[1]{\langle #1|}
\newcommand{\ket}[1]{|#1\rangle}
\newcommand{\braket}[2]{\langle #1|#2\rangle}

\definecolor{ao}{rgb}{0.0, 0.5, 0.0}

\definecolor{gb}{rgb}{0.0, 0.5, 0.5}

\definecolor{rgb}{rgb}{0.5, 0.5, 0.5}

\begin{document}


\title{Quantum Simulation of Abelian Lattice Gauge Theories via State-Dependent Hopping}


\author{A.~S.~Dehkharghani}
\affiliation{Department of Physics and Astronomy, Aarhus University, DK-8000 Aarhus C, Denmark}

\author{E.~Rico}
\affiliation{IKERBASQUE, Basque Foundation for Science, Maria Diaz de Haro 3, E-48013 Bilbao, Spain}
\affiliation{Department of Physical Chemistry, University of the Basque Country UPV/EHU, apartado 644, E-48080 Bilbao, Spain}

\author{N.~T.~Zinner}
\affiliation{Department of Physics and Astronomy, Aarhus University, DK-8000 Aarhus C, Denmark}

\author{A.~Negretti}
\affiliation{Zentrum f\"ur Optische Quantentechnologien and The Hamburg Centre for Ultrafast Imaging, Universit\"at Hamburg, Luruper Chaussee 149, D-22761 Hamburg, Germany}

\date{\today}


\begin{abstract}
We develop a quantum simulator architecture that is suitable for the simulation of $U(1)$ Abelian gauge theories such as quantum electrodynamics. Our approach relies on the ability to control the hopping of a particle through a barrier by means of the internal quantum states of a neutral or charged impurity-particle sitting at the barrier. This scheme is experimentally feasible, as the correlated hopping does not require fine-tuning of the intra- and inter-species interactions. We investigate the applicability of the scheme in a double well potential, which is the basic building block of the simulator, both at the single-particle and the many-body mean-field level. Moreover, we evaluate its performance for different particle interactions and trapping, and, specifically for atom-ion systems, in the presence of micro-motion.
\end{abstract}

\maketitle


\section{Introduction}

Quantum simulation aims at developing devices that can emulate physical systems and experimental conditions that are still poorly understood (e.g., high-temperature superconductivity and quark confinement). It was Feynman that originally suggested to use single purpose quantum computers to simulate a quantum system of interest, which is hardly controllable~\cite{Feynman:1982}. In this regard, trapped ions and quantum degenerate atomic gases are ideal candidates, particularly because of the excellent controllability of the relevant system parameters, e.g., inter-particle interactions and coupling to external fields. Indeed, nowadays atomic quantum simulation of various lattice models relevant in condensed-matter physics is a well-established research area~\cite{Bloch:2012,Blatt:2012}. A new frontier is the quantum simulation of field theories~\cite{Casanova:2011b,Garcia:2015} and strongly correlated systems with dynamical gauge fields, e.g., lattice gauge theories, which are non-perturbative formulations of high-energy physics models such as quantum electrodynamics (QED). Numerical methods that simulate these models range from traditional quantum Monte-Carlo algorithms~\cite{Wilson:1974,Kogut:1983} to tensor networks~\cite{Tagliacozzo2013160,Haegeman:2010,Rico:2014,Tagliacozzo:2014,Banuls:2015,Pichler:2016}. There are problems, however, such as the real-time simulation of heavy ion collisions, that remain intractable with such methods. Hence, simulating such problems with quantum systems specifically tailored for this purpose is necessary. In this regard, let us note that very recently an experiment with a small trapped ion quantum computer has simulated the Schwinger pair production mechanism~\cite{Martinez:2016}, thus demonstrating that such quantum simulations are within reach.

\begin{figure}[ht!]
\includegraphics*[scale=0.29]{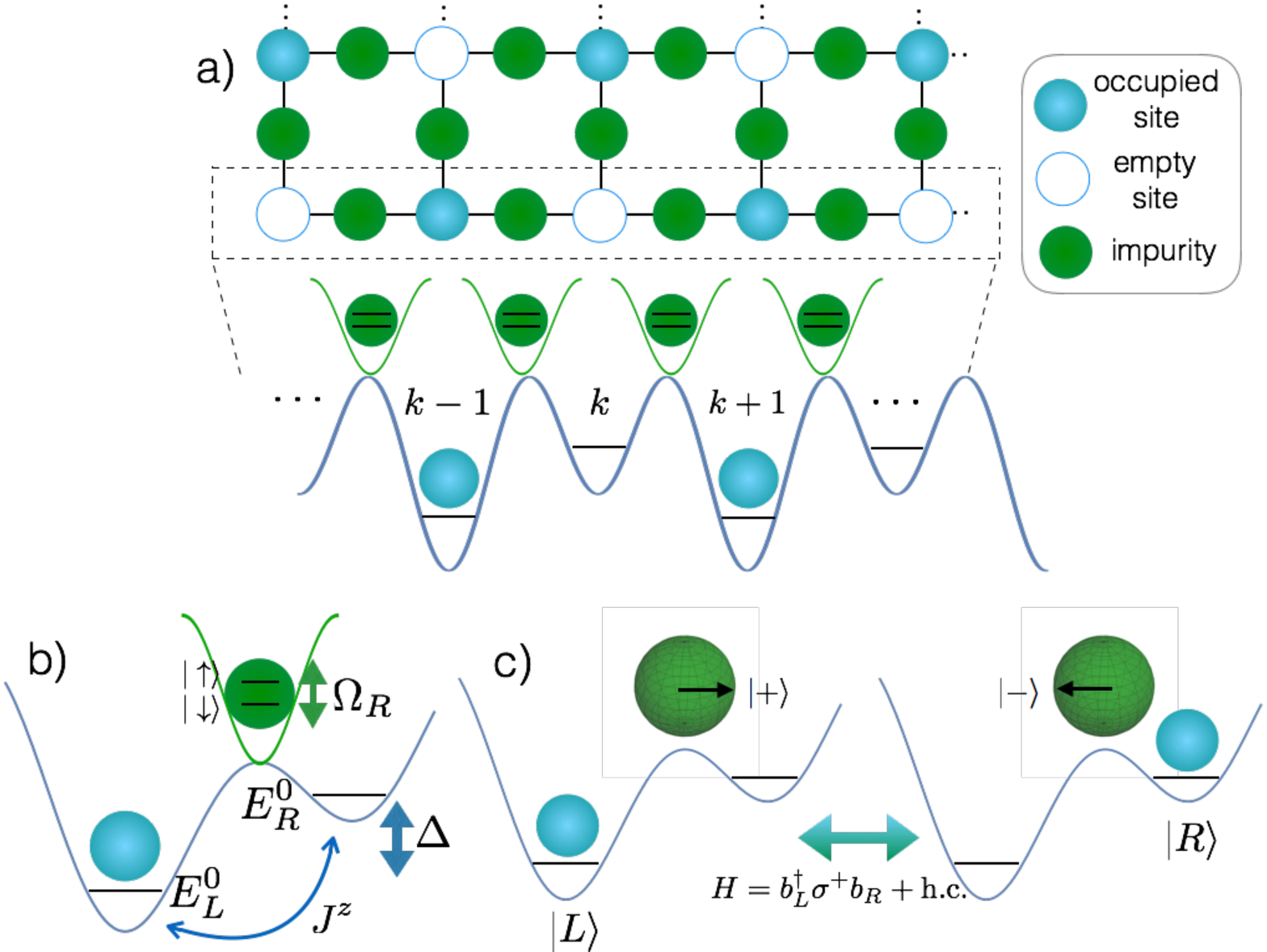}
\caption[]{(Colour online). a) Schematic illustration of the quantum simulator, where atomic particles (blue spheres) are trapped in a superlattice potential (labelled by the lattice sites $\cdots, k-1 , k , k+1, \cdots$), whereas neutral or charged impurities (green spheres) are trapped in another periodic external potential or form an ion crystal, respectively (between the lattice sites, i.e. links). b) Sketch of the double well potential for the particle and the impurity with two internal states $\ket{\uparrow}$ and $\ket{\downarrow}$. c) Sketch of the correlated hopping (left and right picture), where the impurity internal state is illustrated with a black arrow on the equatorial plane of the corresponding Bloch (green) sphere. The offset energy $\Delta$ is chosen to be equal to $\hbar\Omega_R$ (see text).}
\label{fig:sketch}
\end{figure}

The degrees of freedom of usual gauge theories are described by matter fields $\hat{b}_{k}$ and gauge fields $\hat{U}_{k,k +1}$, where $k$ points to a vertex in a regular lattice, while for the sake of simplicity, we refer to $k+1$ its nearest neighbour vertex (see also Fig.~\ref{fig:sketch}). The former fields can be, for instance, fermionic, if we are characterising quark matter, or bosonic, if we are interested in Higgs physics~\cite{Higgs1,Edmonds:2013,Higgs2,Higgs3}. The dynamics of matter and gauge fields is invariant under local gauge transformations and, assuming the locality of the interactions, the most relevant term is $\hat{b}^{\dag}_{k} \hat{U}_{k,k+1} \hat{b}_{k+1}$ + H.c., i.e., a correlated hopping of the matter field mediated by the excitation of the gauge field \cite{HORN1981149,ORLAND1990647,CHANDRASEKHARAN1997455} (e.g., in QED $\hat{U}_{k,k+1}$ is the exponential of the vector potential~\cite{kogut2003phases}). 

Up to now, there are two strategies to build an analog lattice gauge quantum simulator: (a) The (local) gauge symmetry is imposed as a constraint, that is, the gauge variant interactions are cancelled by a large energy penalty~\cite{Banerjee:2012,Marcos:2013} in such a way that the unphysical gauge variant states are very unlikely populated; (b) the local symmetry is mapped into a fundamental symmetry of the system~\cite{Zohar:2016}. In the first strategy, the symmetry is not a fundamental one for the simulator as in (b), but it arises in the low-energy sector of the simulated Hamiltonian such that gauge variant interactions are cancelled by a large energy penalty. For instance, in the proposal of Ref.~\cite{Banerjee:2012} for the quantum simulation of QED, the boson-boson and fermion-boson interactions are chosen to be of the same order such that all gauge variant states are removed from the low-energy sector of the system Hilbert space, and thus fulfilling gauge invariance. Here, we aim at realising a similar gauge invariant Hamiltonian by means of the global symmetry, i.e., strategy (b), which offers interesting advantages, as we will show in the following.
 
To this end, in the next section we introduce the basic formulation of a lattice gauge theory that we use, the so-called Quantum Link Models and we describe how the Higgs or Meissner effect would appear in this context. Then in Sec. III, we summarise the main ingredients of the atomic physics setup needed for the quantum simulation of this type of models via state-dependent hopping. To provide a fine-grained characterisation of the experimental architecture, we develop the microscopic description of the main interactions between the atoms (i.e., the matter field) and the impurities (i.e., the gauge field) that give as a result a state-dependent hopping. We analyse several cases: from a simplified model with a four-level system, through the neutral atom impurity situation, ending at the ion impurity setup. In Sec. V we discuss a many-body problem where the proposed quantum simulator could be an efficient and useful tool, we follow the time evolution of a quenched experiment in lattice gauge theories. Finally, in Sec. VI we discuss parameters for experimental implementations, while in Sec. VII we summarise our findings and provide an outlook. Technical derivations are provided in an appendix.

\section{Quantum link formulation of a $U(1)$ lattice gauge model}

In a $U(1)$ lattice gauge model, the hopping of the matter field is mediated by the gauge field via the term, $\hat b^{\dagger}_{k} \hat U_{k,k+1} \hat b_{k+1}$, which represents the minimal coupling of the matter and gauge degrees of freedom. The fundamental gauge degrees of freedom $\hat U_{k,k+1}$ describe quantum operators associated with the lattice links. The complete Hamiltonian, up to the magnetic term, is given by~\cite{kogut2003phases}

\begin{equation}
\begin{split}
\label{eq:Hqlm}
\hat H_{\text{QLM}}=&-J \sum_{k} \left( \hat b^{\dagger}_{k} \hat U_{k,k+1} \hat b_{k+1} + \text{h.c.} \right) \\
&+ m \sum_{k} \left( -1 \right)^{k} \hat b^{\dagger}_{k} \hat b_{k} + \frac{g^{2}}{2} \sum_{k} \hat E_{k,k+1}^{2}.
\end{split}
\end{equation} 

Here $\hat E_{k,k+1}$ is the electric field operator, $g$ the gauge coupling, $m$ the staggered mass and $J$ the hopping parameter. In the Hamiltonian, the first term corresponds to hopping of the matter field between two adjacent lattice sites, which is associated with a change in the gauge field when the charge moves sites. The second term in the equation, the mass term, defines the staggered matter field, where excitations of these sites corresponds to the creation of a particle-antiparticle pair with a mass gap of $m$. The last term corresponds to the electric field energy.

The electric field  and the link operator are conjugate operators, which obey $[\hat E_{k,k+1} , \hat U_{k,k+1} ] = \hat U_{k,k+1}$. Quantum link models \cite{HORN1981149,ORLAND1990647,CHANDRASEKHARAN1997455} is a framework for lattice gauge theories where the dynamical gauge fields are represented by systems of discrete quantum degrees of freedom with only a finite-dimensional Hilbert space per link. The previous commutation relation is realised by a quantum link operator $\hat U_{k,k+1} = \hat S^{+}_{k,k+1}$, which is a raising operator for the electric flux $\hat E_{k,k+1} = \hat S^{z}_{k,k+1}$ associated with the link connecting neighbouring lattice sites $k$ and $k+1$. Hence, the local gauge Hilbert space is spanned by just $(2S+1)$ quantum states per link. This is in contrast with the original Hamiltonian formulation~\cite{kogut2003phases}, where $\hat{U}_{k,k+1}$ is defined on an infinite dimensional Hilbert space.

A $U(1)$ gauge invariant model is one that has as a local symmetry the conservation of the electric flux around any vertex in the lattice. In any spatial dimension $D$, this constraint reads $\sum^{D}_{\hat{\mu}=1}\left( \hat{E}_{k,k+\hat{\mu}} - \hat{E}_{k-\hat{\mu},k} \right) = \hat{\rho}_{k}$, which means that the difference between the electric fluxes that enter in a vertex minus the electric fluxes that go out from the vertex are equal the charge at this vertex, where $\hat\mu$ is the unit vector in the positive $\mu$ direction. This is nothing but the Gauss law of QED that in the continuum is given by $\vec{\nabla} \vec{E}(x) = \rho(x)$. The generator of the $U(1)$ transformation 

\begin{equation}
\label{generalG}
\hat{G}_{k} = \sum^{D}_{\hat{\mu}=1} \left( \hat{E}_{k,k+\hat{\mu}} - \hat{E}_{k-\hat{\mu},k} \right)- \hat{\rho}_{k}
\end{equation} 
is a constant of motion and, as a symmetry generator, commutes with the Hamiltonian, i.e. $[ \hat H ,  \hat{G}_{k}]=0, ~ \forall k$. Once we have defined the generator of the local $U(1)$ transformation, it is straightforward to check the action of the unitary operator on the relevant fields, for instance,
\begin{equation*}
\begin{split}
e^{i \sum_{k'} \phi_{k'} \hat G_{k'}} \hat b_{k} e^{-i \sum_{k''} \phi_{k''} \hat G_{k''}} &=e^{i\phi_{k}} \hat b_{k} \\
e^{i \sum_{k'} \phi_{k'} \hat G_{k'}} \hat U_{k,k+1} e^{-i \sum_{k''} \phi_{k''} \hat G_{k''}} &=e^{i\phi_{k}} \hat U_{k,k+1} e^{-i\phi_{k+1}} \\
e^{i \sum_{k'} \phi_{k'} \hat G_{k'}} \hat E_{k,k+1} e^{-i \sum_{k''} \phi_{k''} \hat G_{k''}} &=0.
\end{split}
\end{equation*}

For the implementation that we describe, we use a spin-$1/2$ representation of the quantum link in a rotated basis, where $\hat E \mapsto \hat\sigma^{x}/2$ and $\hat U \mapsto \tilde \sigma^{+} = |+\rangle \langle -|$ with $\vert\pm\rangle$ being the eigenstates of the Pauli matrix $\hat\sigma_x$, so the interaction term between the matter and gauge degrees of freedom will be of the form, $\hat b^{\dagger}_{k}  \tilde \sigma^{+}_{k,k+1} \hat b_{k+1}$, and the electric field energy $\hat E^{2} \propto \left[\hat\sigma^{x}\right]^{2}=\mathbb{I}$. In this regard, let us note that in actual QED the electric field and the gauge degrees of freedom are indeed continuum variables. However, in recent numerical studies~\cite{Kuhn:2014,Buyens:2017}, it has been shown that already at small dimensions (on the order of 5), the continuum limit can be reproduced with good accuracy. Hence, by choosing appropriately a manifold with five Zeeman or hyperfine levels, as shown recently in Ref.~\cite{Kasper:2017}, it is possible, also with our scheme, to quantum simulate the Schwinger model in the continuum.  


\subsection{Bosonic Josephson junctions: A toy model for the Higgs physics} 

Due to the bosonic nature of the matter degrees of freedom in the previous $U(1)$ quantum link model, an interesting scenario appears when there is an ensemble of interacting bosonic atoms at every lattice site, for previous implementations of this scenario see Refs.~\cite{Higgs1,Edmonds:2013,Higgs2,Higgs3}. For a large number of bosons per site, we can assume $b_{k} \sim \sqrt{n_{k}} e^{i\theta_{k}}$ with $n_{k}$ the average number of particles within each condensate, which can be fixed to a constant $\frac{N}{2}$, and $\theta_{k}$ the condensate phase. Thus, the first term in Eq. (\ref{eq:Hqlm}) becomes (see also Appendix B for details):

\begin{equation}
\label{eq:BJJ}
\hat H =- J \frac{N}{2} \sum_{k} \left( e^{-i \theta_{k}}\tilde \sigma^{+}_{k,k+1} e^{i \theta_{k+1}} + \text{H.c.} \right).
\end{equation}
This Hamiltonian describes the dynamics of a Bose-condensed matter field interacting with a $U(1)$ gauge field. In such a situation, due to a combination of the spontaneous symmetry breaking of the bosonic field and the gauge interaction, the gauge field acquires a mass, as we will see in the following. 

Let us underscore, however, that contrarily to the large particle number of the gauge field discussed in Ref.~\cite{Kasper:2017}, where the goal is to attain the continuum limit of the gauge field, here we aim at considering large ensembles of particles and weak interactions in the matter field. In such a way the quantum simulator that we are going to introduce in the next sections would in principle enable the investigation of different high-energy and condensed matter physics problems. Indeed, the condensate wave function would represent the bosonic Higgs-field, and thus the Higgs model in high-energy physics or equivalently the Meissner effect in superconductors \cite{fradkin2013field} could we explored.

To be specific, a particular way to obtain a gauge theory with non-zero mass is via the so-called Higgs mechanism, a combination of spontaneous symmetry breaking and gauge symmetry, where the interaction of the Goldstone boson and the gauge boson give rise to a massive theory, i.e., short-range correlated. A simplified model of the spontaneous symmetry breaking is given by the Hamiltonian: $\hat H = -J \sum_{k} \hat b^{\dagger}_{k+1} \hat b_{k} + V\left( \hat \rho_{k} \right) + \text{h.c.}$, where $\hat b_{k}$ are usual bosonic operators at site $k$ and $\hat \rho_{k} = \hat b^{\dagger}_{k} \hat b_{k}$. In our setting the density-dependent potential is given by $V\left( \hat \rho_{k} \right) = \sum_k \epsilon_k \hat\rho_k+U/2\sum_k\hat\rho_k(\hat\rho_k - 1)$, where the first term represents the energy due to the confinement and the second one the interaction between the bosons. If there is a non-trivial potential that depends only on the density of bosons $V\left( \rho_{k} \right)$, as in our physical setup, we can expand the bosonic operators around the potential minima, i.e., $\frac{\partial}{\partial \rho} V \big|_{\rho=\rho_{0}} = 0$ and $b_{k} \sim \sqrt{\rho_{0}} e^{i \theta_{k}}$. Such approximation holds definitely in the large particle limit and for weak inter-particle interactions. Under these assumptions, the Hamiltonian is recast into: $\hat H=2 \rho_{0} \sum_{k} \cos{\left( \theta_{k+1} - \theta_{k} \right)} \sim \int dx \left[ \frac{\partial}{\partial x} \theta (x) \right]^{\dagger} \frac{\partial}{\partial x} \theta (x)$. Here, we assumed that the potential $V$ is around its minimum value and, in the last equality of the Hamiltonian, we assumed that the lattice spacing is small enough as well as the function $\theta (x)$ is continuous and differentiable. When the ground state of the system gets a non-trivial density expectation value, a massless (Goldstone) boson appears due to the spontaneous global symmetry breaking.

Now, if the bosonic degrees of freedom are coupled to a gauge field we have: $\hat H = \sum_{k}J_k \hat b^{\dagger}_{k+1} e^{i g a \hat A_{k+1,k}} \hat b_{k} + V\left(\hat \rho_{k} \right) + \text{h.c.}$, where we assume again that the potential $V(\hat \rho_k)$ depends on the boson density only, $a$ is the lattice spacing and $g$ the coupling constant. Before expanding the interacting Hamiltonian, we perform a local gauge transformation such that: $\hat{\tilde b}_{k} = \hat b_{k} e^{-i \theta_{k}}$ and $ga \tilde A_{k+1,k} = ga \hat A_{k+1,k} + \theta_{k} - \theta_{k+1}$, then the interacting Hamiltonian reduces to a mass term in the gauge field, $\hat H= g^{2} \rho_{0}\int dx ~ \hat A(x)^{2}$. This mechanism is known as Higgs effect in particle physics and Meissner effect in condensed matter physics, and it could be quantum simulated with our proposed implementaion.


\section{System, Target Hamiltonian and Conditions for quantum simulation} 

The system under investigation is illustrated schematically in Fig.~\ref{fig:sketch} a). A superlattice (blue line) with a periodic sequence of double well-like potentials, each trapping neutral atoms (hereafter called particles), is formed. In another periodic potential (green line), which does not influence the former superlattice, either neutral or charged atoms (hereafter called impurities) with two addressable internal states $\ket{\uparrow},\,\ket{\downarrow}$ are trapped at the position of each double well potential barrier. The particles represent the matter field, while the impurities play the role of the dynamical gauge field. In the limit where the dynamics of the impurities is frozen (i.e., tight trapping), the particle-impurity interaction can be treated as an external potential for the particles. As shown in Refs.~\cite{Micheli:2004,Gerritsma:2012,Schurer:2016}, the internal state of the impurity can control the hopping, $J$, of a particle confined, e.g., in a double well potential, via the particle-impurity interaction, and therefore be promoted to an operator $\hat{J}= J^{\downarrow} \ket{\downarrow} \langle \downarrow |+J^{\uparrow}| \uparrow\rangle \langle \uparrow|$. We note that the physical origin of such state dependence is due to the reliance of the particle-impurity scattering length on the electronic internal states, especially at the short-range part of the interaction potential when the particle and impurity electronic clouds do overlap in a collision. Such controlled hopping can be generalised to a lattice, as depicted in Fig.~\ref{fig:sketch} a), in such a way that, within the tight binding approximation, the following Hamiltonian for the particles holds

\begin{align}
\label{eq:BHsigma}
\hat H = \sum_{k} \hat b^{\dag}_{k}   \hat J_{k,k+1} \hat b_{k+1} + \text{H.c.} +\sum_{k} \frac{\hat U_{k}}{2}\hat n_{k}\,(\hat n_{k} -1).
\end{align}
Here $\hat N = \sum_{k}\hat n_{k} = \sum_{k}\hat b^{\dag}_{k}\hat b_{k}$ is the particle number,  $\hat J_{k,k+1}$ is the state-dependent hopping, $\hat{U}_{k}$ denotes the on-site energy, whose state-dependence is very weak~\cite{Gerritsma:2012,Negretti:2014}.

Our goal is to show that the first sum in Eq.~(\ref{eq:BHsigma}) can be mapped into $\propto \sum_{k}  \hat b^\dag_{k}\,   \tilde\sigma^+_{k,k+1}  \hat b_{k+1} +$ H.c. with $\tilde\sigma^+ = |+\rangle \langle -|$, where the eigenstates of the Pauli matrix $\hat\sigma^x$ are given by $\vert \pm\rangle = (|\uparrow\rangle\pm |\downarrow\rangle)/\sqrt{2}$, while the ones of $\hat\sigma^z$ by $|\uparrow\rangle$ and $|\downarrow\rangle$, respectively. Hence, if a particle tunnels from the site indexed by $k$ to the site indexed by $k+1$, then the internal state of the impurity is flipped [see also panel c) of Fig.~\ref{fig:sketch}]. This interaction is gauge-invariant and of the form $\hat{b}^{\dag}_{k}\, \hat{U}_{k,k+1} \hat{b}_{k+1}$ + H.c. 

To demonstrate this, let us for the moment ignore the on-site terms and focus on a single double well only. We start from a microscopic Hamiltonian $\hat H_{(k,k+1)} = \hat H^0_{(k,k+1)} + \hat H^{\text{hopp}}_{(k,k+1)}$ that is split in a local part and an interactive part, 
where

\begin{equation}
\hat H^0_{(k,k+1)} = E_{k}^0 \hat n_{k} + E_{k+1}^0 \hat n_{k+1} + \frac{\hbar \Omega_R}{2} \hat \sigma_{k,k+1}^{x}.
\end{equation}
We note that with such a notation the entire Hamiltonian on the lattice reads

\begin{equation}
\hat H =\sum_k \hat H_{(k,k+1)}+\text{on-site terms}.
\end{equation}
The natural energy scales of each mode are $E_{k}^0$ for the particle at the site indexed by $k$ (i.e., the lowest trapping energy of the particle in the $k$-th site), whereas $\Omega_R$ is the Rabi frequency of the coupling between the impurity internal states (see also Fig.~\ref{fig:sketch}). The hopping part of Eq.~(\ref{eq:BHsigma}) can be rewritten as 

\begin{align}
\hat H^{\text{hopp}}_{(k,k+1)}= \hat b^\dag_{k}\, (J_{k,k+1}^{0}  + J_{k,k+1}^{z} \hat \sigma_{k,k+1}^z) \hat b_{k+1} + \text{H.c.},
\end{align}
where $J_{k,k+1}^{0}=(J^{\uparrow}_{k,k+1}+J^{\downarrow}_{k,k+1})/2$ and $J_{k,k+1}^{z}=(J^{\uparrow}_{k,k+1} - J^{\downarrow}_{k,k+1})/2$. Moving to the interaction picture we obtain:

\begin{equation*}
\begin{split}
\hat H_{\text{rot}} =&  e^{i \hat H^{0}_{(k,k+1)} t} \hat H^{\text{hopp}}_{(k,k+1)} e^{-i \hat H^{0}_{(k,k+1)} t} \\
=& e^{i \left(  E_{k}^0 - E_{k+1}^0 \right) t} \hat b^\dag_{k} \, \hat b_{k+1}  \\
&\left(\frac{J_{k,k+1}^{0}}{2}  + J_{k,k+1}^{z}  e^{-i \hbar \Omega_R t} \tilde\sigma^+_{k,k+1} + \text{H.c.} \right)  + \text{H.c.}
\end{split}
\end{equation*}
By assuming resonant coupling $\Delta =  E_{k}^0 - E_{k+1}^0  = \hbar \Omega_R$ and by applying the rotating-wave approximation, $\{ \vert \Delta \vert, \, \hbar \Omega_R\} \gg \{  J_{k,k+1}^{0} , \, J_{k,k+1}^{z} \}$, we arrive at the desired target  Hamiltonian
\begin{equation}
\hat H^{\text{t}}_{(k,k+1)} = J_{k,k+1}^{z} \hat b^\dag_{k}  \,\tilde\sigma^+_{k,k+1} \hat b_{k+1} + \text{H.c.} 
\end{equation}
Gauge variant term are negligible in the limit where the rotating-wave approximation is valid, for instance, terms like $J_{k,k+1}^{0} e^{i  \Delta  t} \hat b^\dag_{k} \, \hat b_{k+1}$ or $J_{k,k+1}^{z} e^{i 2 \Delta t}  \hat b^\dag_{k} \, \tilde\sigma^-_{k,k+1} \hat b_{k+1} $ oscillate with a high frequency $\Delta$ compared with the energy scales $J_{k,k+1}^{0}$ or $J_{k,k+1}^{z}$, which give a negligible average contribution in a long-term limit. Finally, let us note that since the onsite interaction is of a density-density type, such interaction does not affect the gauge invariance; and in addition, its energy scale and state dependence are typically small~\cite{Gerritsma:2012,Negretti:2014}.

In order to prove the quantum simulation of our scheme, two features have to be checked: (i) the robustness of the local gauge symmetry, and (ii) the correlated hopping of the impurity and the particle modes. To prove the former, we have to evaluate the local generator of the gauge transformation. Being a symmetry, such a generator commutes with $\hat H^{\text{t}}_{(k,k+1)}$. From the general expression Eq. (\ref{generalG}), it can be easily verified for \emph{a single link} $(k,k+1)$ that the operators $\hat G_{k} = \hat n_{k} - \hat\sigma^{x}_{k,k+1}/2$ and $\hat G_{k+1}=\hat n_{k+1} + \hat\sigma^{x}_{k,k+1}/2$ are constants of motion and local generators of the $U(1)$ (phase) transformation of the local operators. Indeed, they are the terms that act non-trivially on the link $(k,k+1)$ from Eq. (\ref{generalG}). We note that in the QED context $\hat\sigma^{x}_{k,k+1}/2$ plays the role of the electric field in-between two sites. Hence, by keeping track of the expectation values of $\hat G_{k}$ and $\hat G_{k+1}$ we can assess the robustness of the gauge invariance, i.e., to which extent the expectation values of these operators evolve in time. Since the effective dynamics of the system corresponds to correlated hopping of the impurity and the particle modes, the feature (ii) can be assessed by looking at correlation functions, for instance, $\mathcal{C}_{k,k+1}(t) = \langle ( \hat b^\dag_{k} \hat b_{k} - \hat b^\dag_{k+1} \hat b_{k+1}) \hat\sigma_{k,k+1}^{x} \rangle - \langle\hat b^\dag_{k} \hat b_{k} - \hat b^\dag_{k+1} \hat b_{k+1}\rangle \langle \hat\sigma_{k,k+1}^{x} \rangle$. Indeed, this observable can distinguish a correlated particle-impurity hopping from the independent particle hopping and impurity internal state flip.


\section{Model system: The double-well}  

In this section, we consider the microscopic description of the proposed quantum simulator, describing the dynamics of the atom and the impurity and the tools to solve the time-dependent Schr\"odinger equation. At the beginning, we analyse a simplified four-level system, where we find all the ingredients needed for the quantum simulation of the abelian lattice gauge model via state-dependent hopping. Then, we consider the full dynamics for both a neutral particle and impurity, where the former is trapped in a double well and the latter in a harmonic trap. Finally, we consider the case of a charged impurity, in particular the case of atom-ion quantum systems, and demonstrate the feasibility of our scheme with such compound system as well.


\subsection{The Hamiltonian}

We shall verify to which extent $\hat H^{\text{t}}_{(k,k+1)}$ can be accomplished by simulating the particle dynamics in a double well interacting with an impurity. We note that the double well represents the basic building block of our quantum simulator (see also Fig.~\ref{fig:sketch}). Specifically, we consider the one-dimensional Hamiltonian $\hat H_{\text{p-i}} = \hat H_\text{p} + \hat H_\text{i} + \hat V_{\text{p-i}} + \hat H_\text{i}^{\text{int}}$, where

\begin{align}
\label{eq:dwH}
 \hat H_{\text{i}}^{\text{int}} &= \frac{\hbar \Omega_R}{2} \hat \sigma_x;
\,\,\, \hat H_{\text{j}}  = -\frac{\hbar^2}{2m_\text{j}}\frac{\partial^2}{\partial x_\text{j}^2} + V_\text{j}^{\mathrm{ext}}(x_\text{s}) \,\,\, \text{j= p, i};\nonumber\\
\hat V_{\text{p-i}} &= \!\!\sum_{s=\uparrow,\downarrow}\upsilon_{\mathrm{1D}}^{e,s}(x_\text{p} - x_\text{i}) + \upsilon_{\mathrm{1D}}^{o,s}(x_\text{p} - x_\text{i}).
\end{align}
Here $V_\text{i}^{\mathrm{ext}}(x_\text{i})=\frac{m_\text{i}\omega_\text{i}^2}{2}x_\text{i}^2$ is the confining potential for the impurity, $V_\text{p}^{\mathrm{ext}}(x_\text{p})= m_\text{p} \omega_L^2(x_{\text{p}}+d_L)^2 [1 - \Theta(x_\text{p})] /2 +[ m_\text{p} \omega_R^2(-x_{\text{p}}+d_R)^2/2+\Delta] \Theta(x_{\text{p}})$ is the double well potential for the bosons with $\Theta(x_\text{p})$ the Heaviside function, $m_\text{p}$ the particle mass, $m_\text{i}$ the impurity mass, and $\Delta$ the energy offset between the two wells (see also Fig.~\ref{fig:sketch}). Furthermore, the particle-impurity interaction is modelled with a contact potential~\footnote{We underscore that such contact potentials are applicable when the effective range of the particle-impurity interaction is smaller than any other length scale involved in the problem.} with even $\upsilon_{\mathrm{1D}}^{e,s}(x) = g_{\mathrm{1D}}^{e,s} \hat\delta_{\pm}(x)$ and odd $\upsilon_{\mathrm{1D}}^{o,s}(x) = g_{\mathrm{1D}}^{o,s} \delta^{\prime}(x)\hat\partial_{\pm}$ terms~\cite{Girardeau:2004,Valiente:2015}, whereas the operators $\hat\delta_{\pm}(x)$ and $\hat\partial_{\pm}$ are defined in the Appendix A. Here $g_{\mathrm{1D}}^{e,s} = -\hbar^2/\mu a_{\mathrm{1D}}^{e,s}$ and $g_{\mathrm{1D}}^{o,s} = -\hbar^2 a_{\mathrm{1D}}^{o,s}/\mu$ with $\mu$ the reduced particle-impurity mass, and $a_{\mathrm{1D}}^{e,s},\,a_{\mathrm{1D}}^{o,s}$ are the 1D spin-dependent scattering lengths for even- and odd-waves, respectively. While the even term has the familiar form of the contact interaction between two ultra-cold bosons, the odd term appears when the interacting particles are spin-polarised fermions or distinguishable particles. As a consequence, the general solution to the scattering problem is a linear combination of even and odd parity solutions. 


\subsection{Solution to the time-dependent Schr\"odinger equation for the double well} 

In general, the time evolution of a state vector is described by the time-dependent Schr\"{o}dinger equation:

\begin{align}
i\hbar\frac{\partial}{\partial t}|\Psi(t)\rangle=\hat H(t) |\Psi(t)\rangle,
\label{eq:SEgeneral}
\end{align}
which has the solution $|\Psi(t)\rangle=\hat U(t)|\Psi(0)\rangle$ with $|\Psi(0)\rangle$ and $\hat U(t)$ being the initial state and the time evolution operator, respectively. In order to solve the corresponding system dynamics practically, we expand the particle-impurity wave function as:
\begin{align}
\Psi(x_{\text p},x_{\text i},s;t) = \sum_{n,m,p} C_{n,m,p}(t) \psi_n(x_{\text p})\phi_m(x_{\text i})\chi_p.
\label{eq:Psitxys}
\end{align}
Here $C_{n,m,p}(t)$ are time-dependent coefficients, $\psi_n(x_{\text p})$ are time-independent basis functions for the particle Hilbert space, $\phi_m(x_{\text i})$ are time-independent basis functions for the impurity Hilbert space, and $\chi_p$ denotes the spinor component of the impurity internal state. By inserting the ansatz~(\ref{eq:Psitxys}) into the Schr\"{o}dinger equation~(\ref{eq:SEgeneral}) and projecting onto the state $\langle \chi_p,\phi_m,\psi_n\vert$, we obtain a set of coupled differential equations for the coefficients $C_{n,m,p}(t)$ that have then to be solved numerically. 


\subsection{A simplified model: The four-level system}

\begin{figure}
\includegraphics[width=0.9\linewidth]{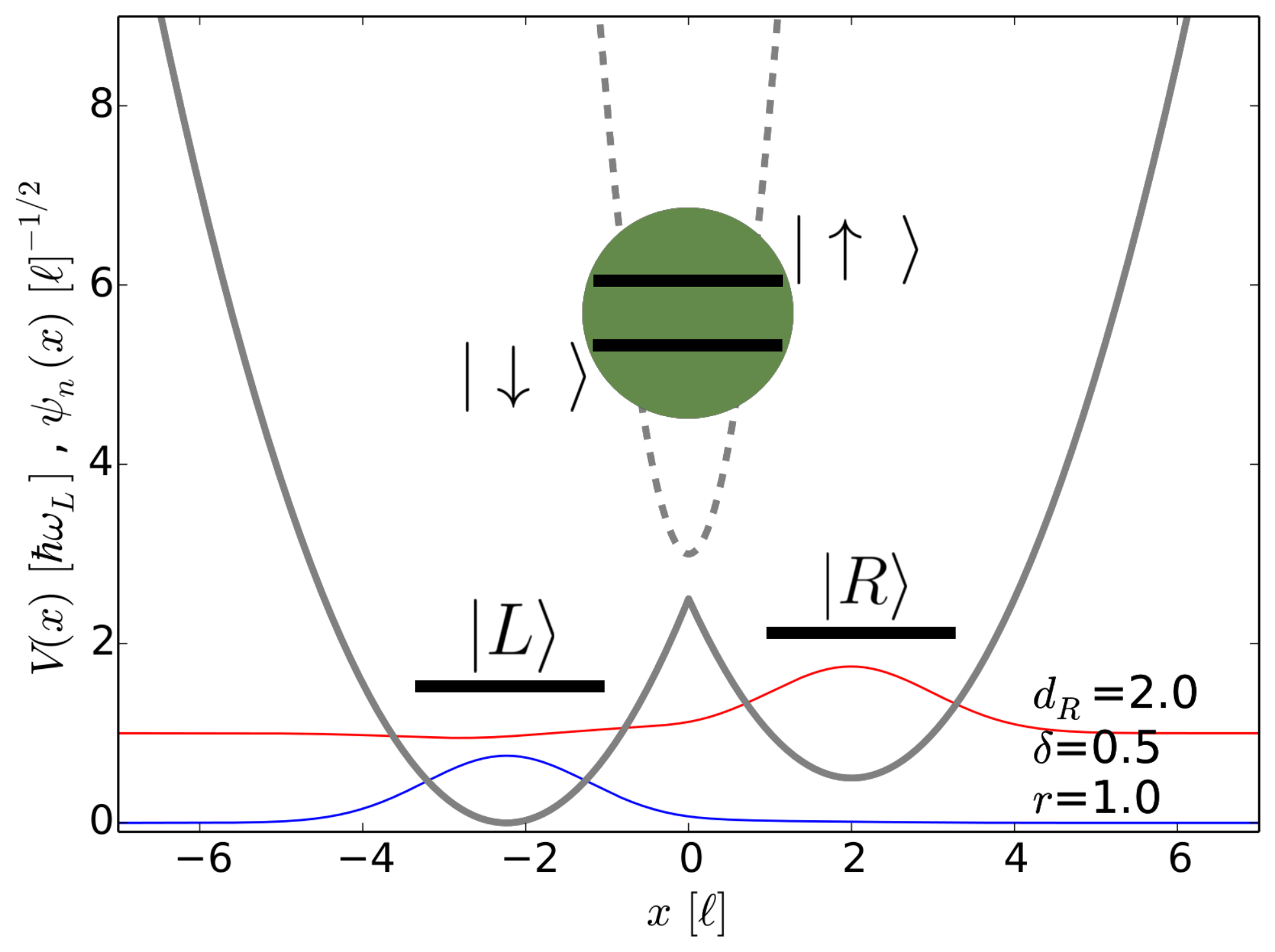}
\caption{Sketch of the four-level particle-impurity system, where the impurity (green circle on the top of the barrier) is treated statically with two internal states, and centred in $x=0$. Here $d_L = -\sqrt{5}\ell$, $d_R = 2.0\ell$, whereas $E^0_{L}/(\hbar\omega_L)=-0.5013$ and $E^0_{R}/(\hbar\omega_L)=-0.0045$ are shifted by $2\hbar\omega_L$ for the sake of clarity. Besides, $\Delta = \delta\hbar\omega_L$ and $\ell=\sqrt{\hbar/(m_{\text p}\omega_L)}$. The left (blue) line and right (red) line represent the two lowest eigenfunctions of the double well potential.}
\label{fig2}
\end{figure}

In the following we consider the case for which the Hamiltonian is time-independent and the impurity is treated statically (i.e., no impurity motion, but only spin dynamics). Thus, the above outlined ansatz reduces to 

\begin{align}
\label{eq:exp-4states}
\Psi(x_{\text p},s;t) = \sum_{n,p} C_{n,p}(t) \psi_n(x_{\text p})\chi_p.
\end{align}
In particular, we shall focus on the special case for which only two motional states of the particle in the double well are considered: One particle-state in the left well (the ground state), $|L\rangle$, one particle-state in the right well (the first excited state of the double well), $|R\rangle$ (see also Fig.~\ref{fig2}). In addition to these states, we have the two impurity spin states $\{|\uparrow\rangle,|\downarrow\rangle\}$ (see also Fig.~\ref{fig2}). Under these assumptions, we have in total four possible state configurations. Moreover, we note that the particle-impurity interaction contributes to the particle Hamiltonian as an external contact potential given by Eq.~(\ref{eq:dwH}) centred in $x_{\text i}=0$, where the impurity is assumed to be located (see also Fig.~\ref{fig2}). Thus, the particle-impurity Hamiltonian is given by 

\begin{align*}
\hat H =& -\frac{\hbar^2}{2m_\text{p}}\frac{\partial^2}{\partial x_{\text p}^2} + V_\text{p}^{\mathrm{ext}}(x_{\text p}) \\
&+ \frac{\hbar \Omega_R}{2} \hat \sigma_x\\ &+\sum_{s=\uparrow,\downarrow}\upsilon_{\mathrm{1D}}^{e,s}(x_\text{p}) + \upsilon_{\mathrm{1D}}^{o,s}(x_\text{p})\\
=& ~\hat H_{\text{p}} + \hat H_{\text{i}}^{\text{int}} + \hat V_{\text{p-i}}.
\end{align*}
Hence, we form the basis $\{|L,\downarrow\rangle, |R,\downarrow\rangle,|L,\uparrow\rangle,|R,\uparrow\rangle\}$, where $\hat H_{\text{p}} |L\rangle= E_L^0 |L\rangle$ and $\hat H_{\text{p}} |R\rangle= E_R^0 |R\rangle$. Hence, in this basis, the particle-impurity Hamiltonian has the following matrix form:

\begin{align}
\begin{bmatrix}
       E^0_{L}+J^{\downarrow}_{LL}& J^{\downarrow}_{LR} & \frac{\hbar\Omega_R}{2}& 0\\[0.3em]
       J^{\downarrow}_{RL} & E^0_{R}+J^{\downarrow}_{RR} & 0 & \frac{\hbar\Omega_R}{2}\\[0.3em]
       \frac{\hbar\Omega_R}{2} & 0 & E^0_{L}+J^{\uparrow}_{LL} & J^{\uparrow}_{LR} \\[0.3em]
       0 & \frac{\hbar\Omega_R}{2} & J^{\uparrow}_{RL} & E^0_{R}+J^{\uparrow}_{RR}\\[0.3em]
     \end{bmatrix},
     \nonumber
\end{align}\\
with 

\begin{align}
J^s_{NM}&=\langle N|\upsilon_{\mathrm{1D}}^{e,s}|M \rangle+\langle N|\upsilon_{\mathrm{1D}}^{o,s}|M \rangle\nonumber\\
&=\langle N|g_{\mathrm{1D}}^{e,s}\hat{\delta}_{\pm}|M \rangle+\langle N| g_{\mathrm{1D}}^{o, s} \delta^{\prime}(x_{\text p})\hat{\partial}_{\pm}|M \rangle\nonumber\\
&=g_{\mathrm{1D}}^{e,s}\cdot N(0)\cdot [M(0^+)+M(0^-)]/2 \nonumber\\
&-g_{\mathrm{1D}}^{o,s}\cdot N^{\prime}(x_{\text p})|_{x_{\text p}=0}\cdot [M^{\prime}(0^+)+M^{\prime}(0^-)]/2
\end{align}
for $s=\uparrow,\downarrow$ and $N,M=L,R$. Here the apex $^{\prime}$ denotes the spatial derivative. Exemplary values of the parameters previously introduced are: $g_{\mathrm{1D}}^{o,\uparrow}=\hbar\omega_L\ell^3$ and $g_{\mathrm{1D}}^{e,\uparrow}=\hbar\omega_L\ell$ imply $J^\uparrow_{LR}=J^\uparrow_{RL}=-0.011\hbar\omega_L$, whereas $g_{\mathrm{1D}}^{o,\downarrow}=-\hbar\omega_L\ell^3$ and $g_{\mathrm{1D}}^{e,\downarrow}=-\hbar\omega_L\ell$ imply $J^\downarrow_{LR}=J^\downarrow_{RL}=-0.011\hbar\omega_L$. Note that if the wells are too separated, then the hopping rates will be exponentially suppressed, and therefore no tunnelling occurs. 

\begin{figure}[t]
\includegraphics[width=\linewidth]{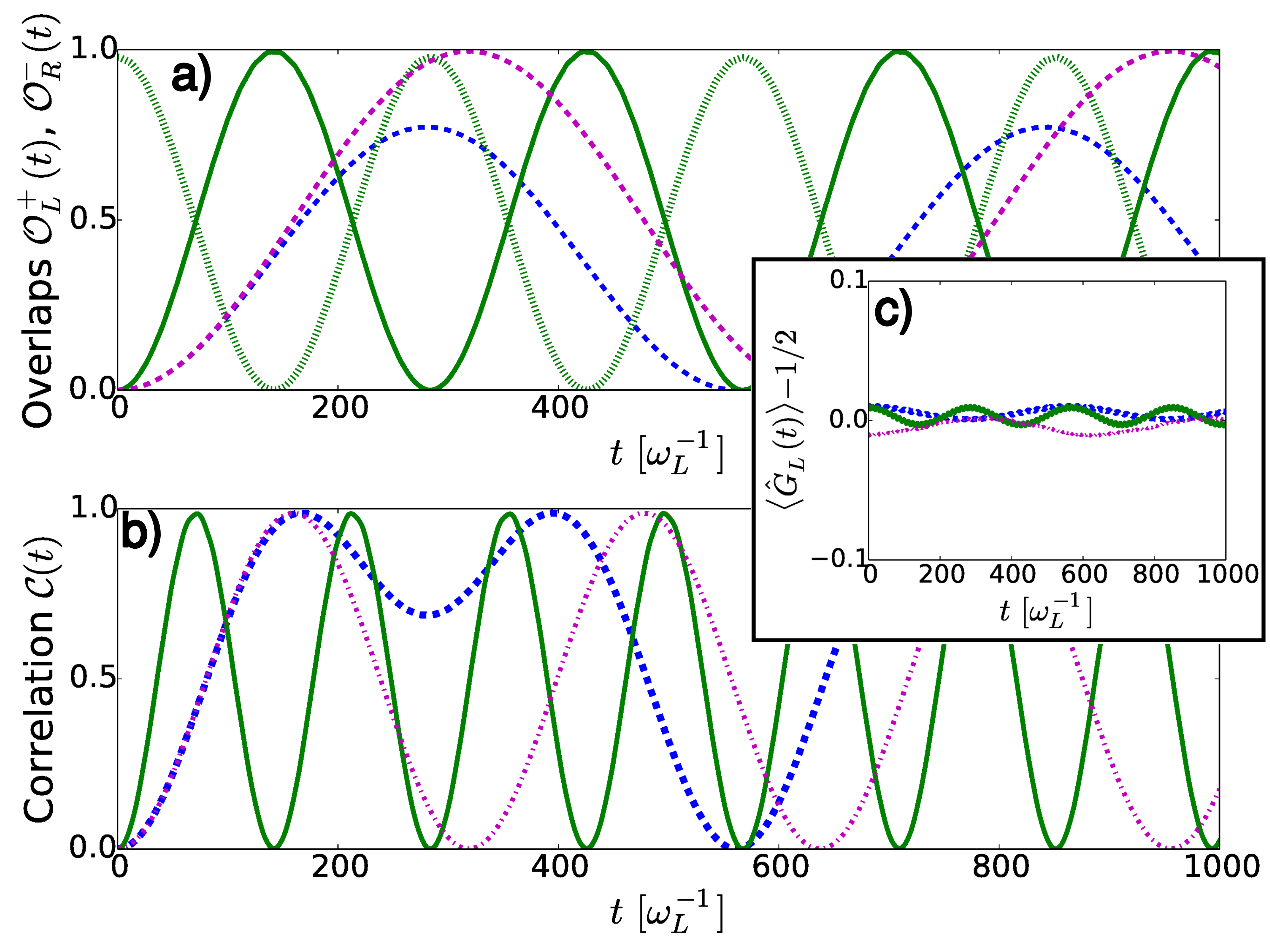}
\caption{(Colour online). Panel a): Overlap $\mathcal{O}_{L}^{+}(t)$ [$\mathcal{O}_{R}^{-}(t)$ is shown only when the coupling constants have the same strength, but opposite sign -- green dashed line]. Panel b): Correlated particle-impurity hopping $\mathcal{C}(t)$. Panel c): Expectation value of the operator $\hat G_L-1/2$ ($\langle\hat G_R\rangle$ is specular to $\langle\hat G_L\rangle$). The lines correspond to three different scenarios: $g_{\mathrm{1D}}^{e,\uparrow} = -g_{\mathrm{1D}}^{e,\downarrow} = \hbar\omega_L \ell$, $g_{\mathrm{1D}}^{o,\uparrow} = -g_{\mathrm{1D}}^{o,\downarrow} = 0.1\,\hbar\omega_L \ell^3$ [solid green $\mathcal{O}_{L}^{+}(t)$ and dashed green $\mathcal{O}_{R}^{-}(t)$ line], $g_{\mathrm{1D}}^{e,\uparrow} = 0.1\cdot g_{\mathrm{1D}}^{e,\downarrow} = \hbar\omega_L \ell$, $g_{\mathrm{1D}}^{o,\uparrow} = 0.1\cdot g_{\mathrm{1D}}^{o,\downarrow} = 0.1\,\hbar\omega_L \ell^3$ (dashed blue lines and dashed-dotted magenta lines). The only difference between the dashed blue and dashed-dotted magenta lines is how they are tuned to resonance. The dashed blue is set to have $\hbar\Omega_R=E^0_R-E^0_L$, whereas the dashed-dotted magenta line is set to have $\hbar\Omega_R=E^0_R-E^0_L+(J^{\downarrow}_{RR}+J^{\uparrow}_{RR}-J^{\downarrow}_{LL}-J^{\uparrow}_{LL})/2$. In all panels $m_{\text p}/m_{\text i} = 1$ and $\omega_{\text{i}} = 10^3\omega_L$. Furthermore, $\omega_L = \omega_R$, $d_L =\sqrt{5}\, \ell$, $d_R = 2 \ell$ for a barrier height of $4\,\hbar\omega_L$.}
\label{supp:fig3}
\end{figure}

Now, we have simulated the corresponding dynamics by assuming that for $t=0$ the initial state is $|\Psi(t = 0)\rangle=\frac{|R,{\uparrow}\rangle-|R,{\downarrow}\rangle}{\sqrt{2}}\equiv|R,-\rangle$. Hence, the only non-vanishing coefficients are: $C_{R,{\downarrow}}=-1/\sqrt{2}$ and $C_{R,{\uparrow}}=1/\sqrt{2}$. By solving the corresponding coupled differential equations for the coefficients $C_{n,p}(t)$, we can extract the probability of finding the particle in the left well with the flipped spin state for the impurity as $\mathcal{O}^+_L(t) = \vert C_{L,\uparrow}(t) + C_{R,\downarrow}(t)\vert^2/2$ [similarly for $\mathcal{O}^-_R(t)$]. Examples of such calculations for different interaction strengths are illustrated in Fig.~\ref{supp:fig3}. As one can see, in the case the couplings for the two spin states have opposite sign, but have the same strength (green lines in Fig.~\ref{supp:fig3}), the particle-impurity state oscillates purely between the states $|R,-\rangle$ and $|L,+\rangle$. This situation represents the ideal scenario we are aiming for in our study. 

Furthermore, the overall picture is maintained if the strengths are different (not shown). On the other hand, if the couplings have the same sign (dashed blue lines), the process is significantly modified with respect to the desired ideal scenario (green solid lines). This is because the resonance condition has to be replaced with $\hbar\Omega_R=E^0_R-E^0_L+(J^{\downarrow}_{RR}+J^{\uparrow}_{RR}-J^{\downarrow}_{LL}-J^{\uparrow}_{LL})/2$. Then, the process works again very well as in the other cases (see dash-dotted magenta lines). Hence, the desired process can take place regardless the values of the coupling constants for the even and odd wave and for the two spin-states of the impurity. Thus, this proves that our schemes does not require fine tuning of the inter-particle interactions.

\subsection{Simulation for neutral impurity-particle systems}
\label{sim-neutral}

\begin{figure}[ht!]
\includegraphics*[width=1\linewidth]{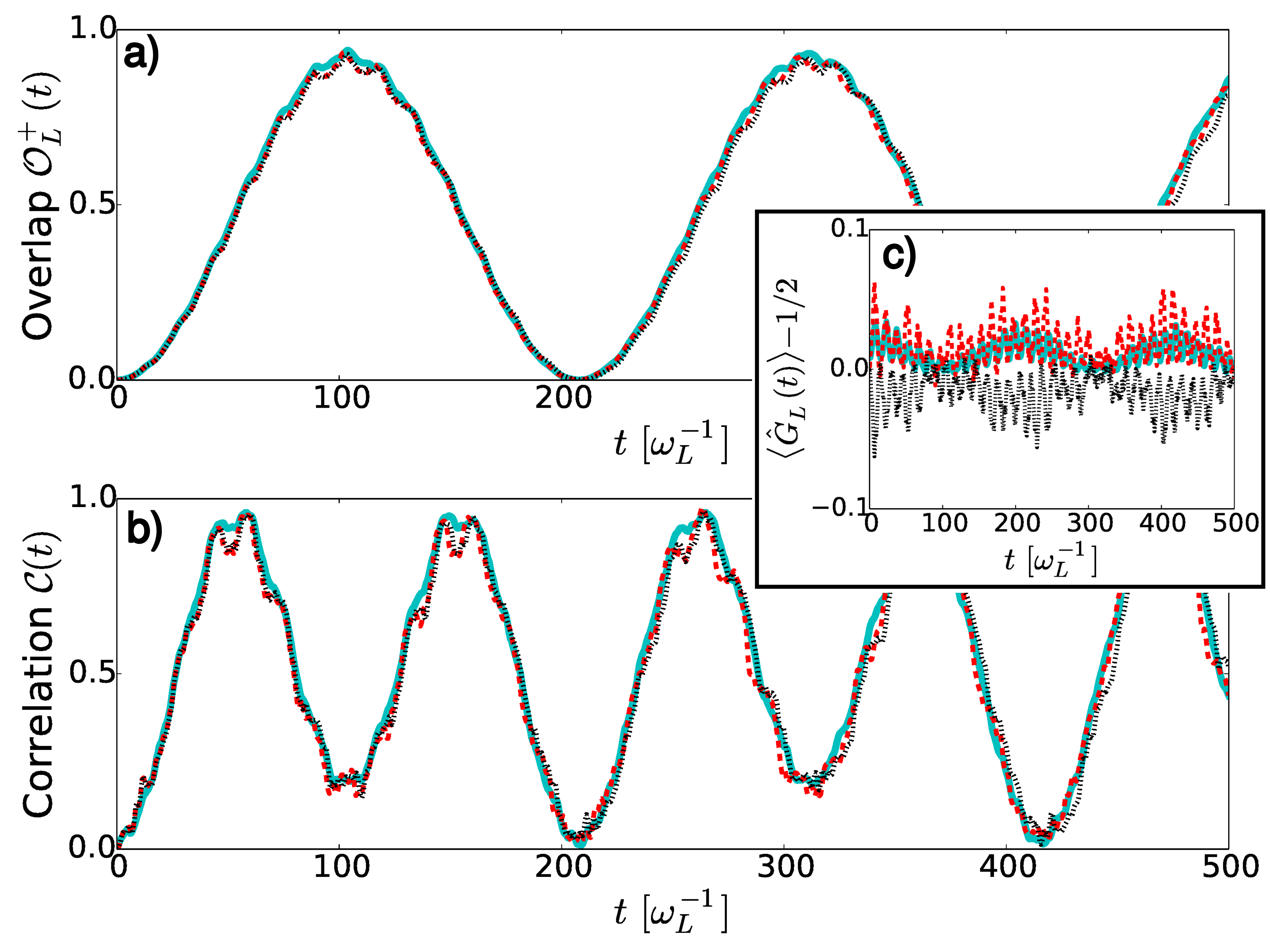}
\caption[]{(Colour online). a) Overlap $\mathcal{O}_{L}^{+}(t)$ b) Correlated particle-impurity hopping $\mathcal{C}(t)$ and c) Expectation value of $\hat G_L - 1/2$ ($\langle\hat G_R\rangle$ is specular to $\langle\hat G_L\rangle$) for three different scenarios: static impurity (solid cyan), moving impurity together with the quench of the barrier height (dashed red), and micro-motion (dotted black). In all panels $m_{\text p}/m_{\text i} = 1$, but $\omega_{\text{i}} = 10\omega_L$ for the moving neutral impurity, and $\omega_{\text{i}} = 100\omega_L$ for the ionic impurity micro-motion with $\Omega_{\text{rf}}/\omega_L = 2500$. The particle-impurity interaction parameters are: $g_{\mathrm{1D}}^{e,\uparrow} = -g_{\mathrm{1D}}^{e,\downarrow} = \hbar\omega_L \ell$, $g_{\mathrm{1D}}^{o,\uparrow} = -g_{\mathrm{1D}}^{o,\downarrow} = 0.1\,\hbar\omega_L \ell^3$, with $\ell=\sqrt{\hbar/(m_{\text{p}}\omega_L)}$. Furthermore, $\omega_L = \omega_R$, $d_L=\sqrt{5}\, \ell$, $d_R = 2 \ell$ for a barrier height of $4\,\hbar\omega_L$, whereas in the quenched case $\omega_R = 1.4\, \omega_L$, $d_L \simeq 2.97 \ell$, $d_R = 2\, \ell$ for a barrier height of $5.92\,\hbar\omega_L$.}
\label{fig:dw}
\end{figure}

Upon the previous findings on the four-level system, we investigate now the full dynamics for a neutral impurity and a particle in the double well potential. We begin by consider a tightly trapped impurity such that $\hat H_\text{i}\equiv 0$ (i.e., static impurity at $x_{\text{i}}=0$). In this case the eigenvalue problem for the particle is analytically solvable, as shown in the Appendix A. We assume a well separation $d_L + d_R$ such that $J_{\uparrow,\downarrow}$ is not too small, but still smaller than $\Delta$ and $ \hbar \Omega_R$. This means that in such a configuration the particle and the impurity are interacting. As initial condition for the time-dependent Schr\"odinger equation we choose the particle state $\ket{R}$, i.e., the first excited eigenstate of $\hat H_\text{p}$, and for the impurity the state $\ket{-}$ (see Fig.~\ref{fig:sketch}). We remind that the goal is to obtain, after a certain time, the target state $\ket{+}$ for the impurity and the particle in the left well, but not necessarily in the ground state $\ket{L}$ of $\hat H_\text{p}$. The result of this analysis is illustrated in Fig.~\ref{fig:dw}, where the overlap $\mathcal{O}_{L}^{+}(t)$ is shown in the upper panel. Here $\mathcal{O}_{y}^{\alpha}(t) =\int_{\mathbb{R}^y}d x_\text{p} \int_{\mathbb{R}}d x_\text{i}\vert\langle x_\text{p},x_\text{i},\alpha_\text{i}\vert\psi_\text{p-i}(t)\rangle\vert^2$ with $\alpha=\pm$, $y=L,R$, $\mathbb{R}^{R,L}\equiv \mathbb{R}^{\pm}$, and where $\psi_\text{p-i}(t)$ is the time evolved particle-impurity state. As the solid lines show, after a time $t_{\max}\approx 104 /\omega_L$ the desired target state is reached [$\mathcal{O}_{L}^+(t_{\max})\simeq 0.94$]. Further, in the lower panel the time evolution of the correlation $\mathcal{C}(t)\equiv\mathcal{C}_{L,R}(t)$ is displayed with revivals in agreement with the evolution of the overlaps. This clearly demonstrates that the particle hopping is correlated with the internal state flip of the impurity. The inset illustrates the expectation value of $\hat G_{L}$ (similarly for $\hat G_{R}$). It shows that the generator of the gauge transformation is, to a good approximation, a constant of motion. Hence, the microscopic dynamics of such a setup corresponds to our target Hamiltonian. In reference to the four-level system studied previously, we see that by including more states in the total wave function expansion~(\ref{eq:exp-4states}), more excited states participate in the dynamics. Thus, the sine-like behaviour of the populations of the four-level scheme is a bit hampered and additional oscillations with smaller amplitudes on the top of the main oscillation appear.

The desired state transfer is also obtained when the static impurity assumption is relaxed (i.e., $\hat H_\text{i}\ne 0$). We obtain essentially the same result as for the static particle for $\omega_{\text{i}} = 10^2\omega_L$ (not shown). We note that this situation can be realised experimentally by starting at $t=0$ with a height of the double well barrier such that no particle-impurity interaction takes place and then suddenly quenching it to a value for which the two systems interact and such that the conditions $\vert \Delta \vert = \hbar \Omega_R$ and $\{ \vert \Delta\vert , \, \hbar \Omega_R\} \gg \{  J^{0} , \, J^{z} \}$ are fulfilled. This is illustrated in Fig.~\ref{fig:dw}, where the desired behaviour is observed. Hence, the impurity trap does not play any major role in the dynamics (e.g., 6\% overlap reduction if $\omega_{\text{i}} = \omega_L$).

Finally, let us note that we have also considered the feasibility of the scheme, particularly of the correlated hopping, in the case the particle system in the double well is replaced by a condensate in contact with a two-level spin impurity. We found that even within mean-field theory for the bosonic ensemble and by means of optimal control techniques such a process is not possible and we attribute this failure mainly to nonlinear effects of inter-particle interactions. A detailed analysis together with the derivation of the corresponding equations of motion is provided in the Appendices C and D.


\subsection{Simulation for the atom-ion system} 

The situations described so far are representatives of an ultra-cold neutral particle-impurity system, such as an atomic boson and fermion. We have also considered the case in which the impurity is replaced by an ion. The system then resembles the ion-controlled double well of Ref.~\cite{Joger:2014}. Here, however, we assume that the long-range atom-ion polarisation potential can be approximated by the contact interaction~(\ref{eq:dwH}), which is reasonable if the separation between the ions is larger than the range of the atom-ion interaction~\cite{Negretti:2014}. Additionally, we have included the effect of the time-dependent radio-frequency (rf) fields, since most of the ion experiments use the so-called Paul-trap~\cite{Leibfried:2003}. In this way we can assess the impact of the micro-motion on the dynamics. More precisely, we have considered the Hamiltonian for an ion in one spatial dimension subjected to static and time-dependent electric fields given by~\cite{Cook:1985}

\begin{align}
\hat H_{\text i}(t)=-\frac{\hbar^2}{2m_{\text i}}\frac{\partial^2}{\partial x_{\text i}^2} + \frac{m_i\Omega^2(t)}{8}x_{\text i}^2
\end{align}
where $\Omega(t)^2=\Omega_{\text{rf}}^2 \cdot [a+2q\cos(\Omega_{\text{rf}} t)]$ with $\Omega_{\rm rf}$ the driving frequency, and $|a|\ll |q|<1$ being some trapping geometrical factors (typical values used in experiments are $|a| \ll |q|\approx0.2$ for a linear Paul trap). For the numerical simulations, however, it is more convenient to recast it as:
\begin{align}
\hat H_{\text i}(t) = -\frac{\hbar^2}{2m_{\text i}}\frac{\partial^2}{\partial x_{\text i}^2}+\frac{m_{\text i}\omega_{\text i}^2}{2}x_{\text i}^2+\frac{m_{\text i}\omega_{\text i}^2}{2}m_{\text i}x_{\text i}^2 \left[\frac{\Omega(t)^2}{4\omega_{\text i}^2}-1\right]
\end{align}
where $\omega_{\text i}=\frac{\Omega_{\text{rf}}}{2}\sqrt{a+q^2/2}$ is the so-called secular frequency, that is, the effective harmonic trap frequency of the ion, if the micro-motion would be neglected. Thus, we chose the basis functions $\phi_m(x_{\text i})$ for the expansion of the particle-ion wave function~(\ref{eq:Psitxys}) as the eigenstates of the harmonic oscillator with frequency $\omega_{\text i}$.

The result is shown in Fig.~\ref{fig:dw} (dotted lines) for a driving frequency $\Omega_{\text{rf}}/\omega_L = 2500$. Similarly to the quenched barrier height case, more pronounced wiggles are observed, but the overall impact is in no way detrimental. Hence, the analysis demonstrates that also with the compound atom-ion system the scheme can be implemented, especially for pairs with a small atom-ion mass ratio (e.g., Li-Yb$^+$), as the influence of the micro-motion~\cite{Cetina:2012,Joger:2014} and the spin-orbit coupling~\cite{Tscherbul:2016} is expected to be small. We note that very recently experimental measurements on charge exchange show that the corresponding rate is significantly smaller than the Langevin collision rate, thus confirming that for Li-Yb$^+$ the required coherence for achieving state-dependent tunnelling is indeed possible~\cite{Joger:2017}.


\section{False vacuum simulation} 

An instance for which the proposed quantum simulator becomes relevant is in out-of-equilibrium dynamics, where the system is initially prepared in a quantum state that is not an eigenstate of the target Hamiltonian, e.g., due to a collision or a sudden quench of the parameters that define the dynamics. The notion of false vacuum decay is linked to out-of-equilibrium physics, in the sense, that the system is prepared in a state (the so-called false vacuum) that is not an eigenstate of the Hamiltonian with which the system evolves.

\begin{figure}[t]
\includegraphics[width=\linewidth]{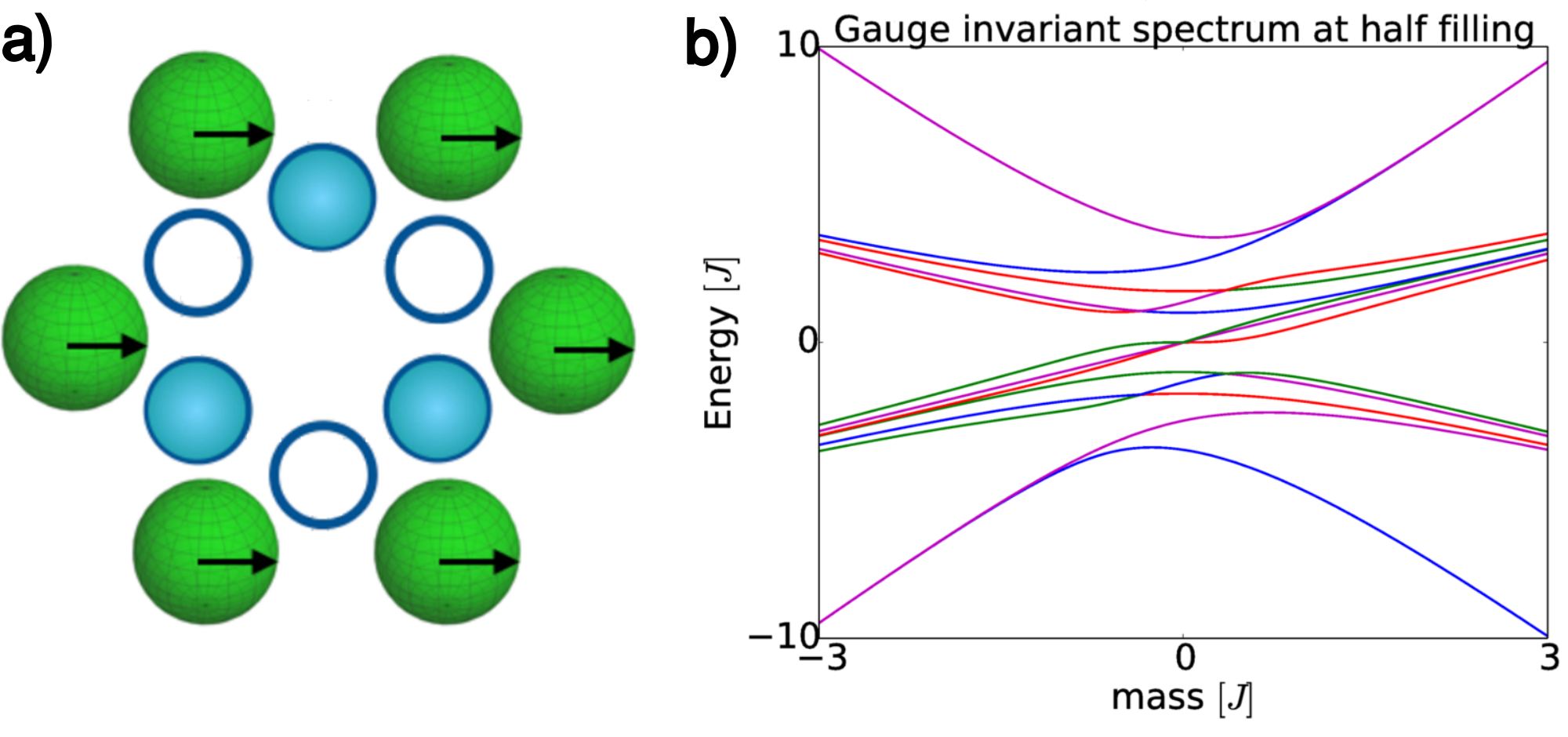}
\caption{a) Minimal instance of 6 sites and 6 impurities, where a demonstration of a lattice gauge simulator could be envisioned. An example of a quenched experiment is explained in the main text. b) Gauge invariant spectrum at half-filling of the proposed model as a function of the mass parameter in the Hamiltonian (\ref{ham idea}). As it can be seen from the degeneracy of the low-energy spectrum, there is a region with a unique ground state and a region with a two-fold degenerate ground state subspace. The numerical calculation has been performed by exact diagonalisation and by assuming periodic boundary conditions.}
\label{suppfv}
\end{figure}

As we have seen, the precise Hamiltonian that characterises the dynamics is given by

\begin{equation}
\label{ham idea}
\hat{H}= -J\sum_{k} \left[ \hat{b}_{k}^{\dagger} \tilde{\sigma}^{+}_{k,k+1} \hat{b}_{k+1} + \text{h.c.} \right] + m \sum_{k} \left( -1 \right)^{k} \hat{b}^{\dagger}_{k} \hat{b}_{k}.
\end{equation}
Such a model describes a bosonic matter field $\hat{b}_{k}$ with a staggered mass $m$ coupled to a gauge field $\tilde{\sigma}^{+}_{k,k+1}= | + \rangle_{k,k+1} \langle - |$. Note that the mass term in Eq. (\ref{ham idea}) is realised in our quantum simulator with the double well potential (or superlattice in a lattice system). Due to the mass term, at half filling, the vacuum of the decoupled model corresponds to a staggered distribution of particles depending on the sign of the mass (i.e., in our simulator, deeper wells correspond to a negative sign, whereas energetically higher wells to a positive sign). At this point, it is worth emphasise the role of the superlattice: In a first place, we have seen that it is needed to implement the desired Hamiltonian in an interacting picture, whereby the difference of the energy between the double wells $\Delta$ matches the local Rabi frequency $\Omega_{R}$ of the impurity that is placed on the links between the sites. Then, if the resonance condition is not completely fulfilled, the detuning between these energy levels takes the role of a staggered mass $m$ that characterises two different type of sites: Particle-sites, when the mass term is positive, $m>0$; antiparticle-sites, when the mass term is negative, $m<0$. Moreover, at half-filling, the sites with negative mass will be filled, defining the reference state of the matter field.

Further, because of the gauge invariance when we consider the gauge field configuration on these staggered particle distributions, there are three possible (ground) states:

\begin{itemize}
\item Positive mass, $(m\gg0)$:

$|g_0\rangle = |\cdots 1_{2k-1} , +_{2k-1,2k}, 0_{2k} , -_{2k,2k+1}, 1_{2k+1} , \cdots \rangle$

\item Negative mass, $(m\ll 0)$:

$|g_+\rangle = |\cdots 0_{2k-1} , +_{2k-1,2k}, 1_{2k} , +_{2k,2k+1}, 0_{2k+1} , \cdots \rangle$

$|g_-\rangle = |\cdots 0_{2k-1} , -_{2k-1,2k}, 1_{2k} , -_{2k,2k+1}, 0_{2k+1} , \cdots \rangle$
\end{itemize}

The first thermodynamical property that characterises all these states is the electric flux density of the system, defined as $\frac{1}{N}\sum_{k} \hat\sigma^{x}_{k}$, i.e., the average value of the configurations of the state of the impurity. The first state $|g_{0}\rangle$ have zero net electric flux density, while the other two has positive, $|g_{+}\rangle$, and negative, $|g_{-}\rangle$, electric flux, respectively.

This cartoon structure of the vacuum of the system remains with the full interacting Hamiltonian~(\ref{ham idea}), i.e., the vacuum of the full Hamiltonian with positive mass is unique with zero electric flux, while with negative mass the ground state manifold is two-fold degenerate with positive and negative value of the electric flux (see Fig.~\ref{suppfv}). These two behaviours, in fact, define two phases separated by a quantum phase transition at a finite negative $m$. The quantum phase transition goes from a charge and parity ordered phase with non-zero electric flux to a disordered one with a net zero electric flux configuration.

\begin{figure}[ht!]
\includegraphics*[width=\linewidth]{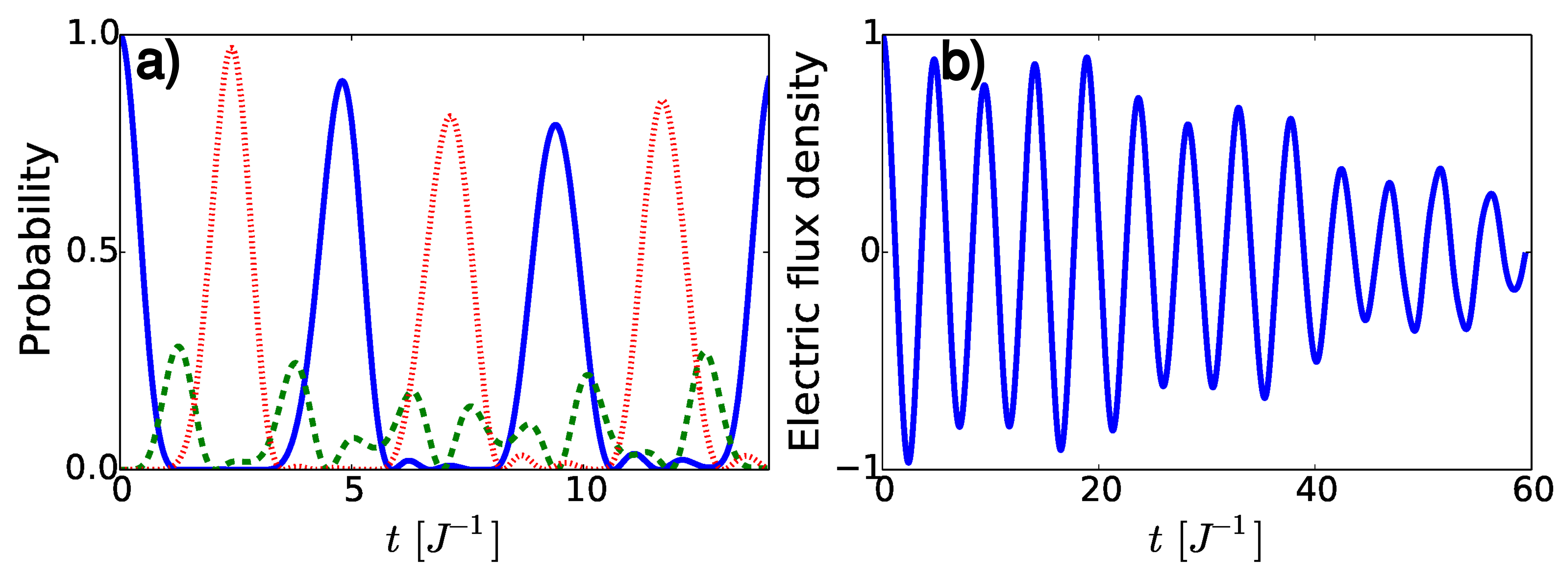}
\caption[]{(Colour online). a) Probabilities of the three states that characterise the different phases of the model: (dashed green) probability of being in a disordered state; (solid blue and dotted red) probability of being in one of the two charge and parity broken states, where the system is prepared in the blue state. b) Evolution in real time of the electric flux density in a quenched experiment, where the system is initialised in the same state as in a), with a non-zero value of the electric flux and it evolves oscillating around a zero value.}
\label{fig:false}
\end{figure}

Once we have identified the different phases and different vacua identified by the charge and parity symmetries \cite{Rico:2014}, we could perform a quenched experiment where, the system is prepared in one of the ground states that break the symmetries and we let it evolve with a Hamiltonian from the disorder phase. Then we expect non-trivial oscillations of the order parameter, i.e., the net electric flux, and oscillations between these states. Figure~\ref{fig:false} illustrates how the electric flux density evolves from a non-zero value to a vanishing one. This type of toy experiments are examples of out-of-equilibrium dynamics, where the initial state of a many-body system is trapped in a local energy minima after a collision or a sudden change of the thermodynamic conditions, and then, the system evolves from the highly excited state.

\section{Experimental considerations} 

Typically, time scales of ultra-cold atom experiments do not exceed a few hundred of milliseconds. This means that in order to attain the desired state in a reasonable time, we need a trap frequency of a few kHz and similarly for the Rabi frequency. For instance, for $\omega_L / (2\pi)=$ 1 kHz we would have $t_{\max}\approx 104/\omega_L \simeq 17$ ms. In the case of a  quasi-1D setting, a transverse frequency for the particle $\sim20\omega_L$ is necessary to keep the transverse motion frozen to its ground state. These frequencies are attainable in optical lattice or atom-chip experiments and coherence times of a few hundred milliseconds are easily obtainable. For the atom-ion system, the condition is even more easily satisfied: For a driving frequency $\Omega_{\text{rf}}/(2\pi) = 2.5$ MHz, we obtain a secular frequency $\omega_{\text i}/(2\pi) \simeq 254$ kHz $\gg\omega_L$. As far as the odd-wave coupling constants $g_{\mathrm{1D}}^{o,\uparrow,\downarrow}$ are concerned, we note that they do not play an essential role for the successful realisation of the process. Since odd-wave contributions at low temperatures are typically small, we used small values for $g_{\mathrm{1D}}^{o,\uparrow,\downarrow}$ and, in principle, they may be negligible on short time-scales. Instead, the even coupling constants $g_{\mathrm{1D}}^{e,\uparrow,\downarrow}$ play a crucial role for the implementation of our scheme. Note, however, that any kind of combination $g_{\mathrm{1D}}^{e,\uparrow,\downarrow}$ will work, a particular choice of such couplings relies essentially on how fast the process has to be with respect to the decoherence times involved in the specific atomic system used. In fact, the correlated hopping vanishes only in the special case where the couplings have the same strength and sign, since here $J_k^z$ becomes zero. We underscore, however, that the proposed scheme can be generalised to higher dimensions and suitable parameters can be determined as well.


\section{Conclusions and outlook} 

In this work we have proposed an alternative route to the quantum simulation of lattice gauge theories by means of compound atomic quantum systems like Bose-Fermi or atom-ion systems. Our approach does not require any precise control of the particle-impurity interactions, and therefore it offers advantages compared to previous proposals~\cite{Banerjee:2012,Kasper:2017}. Furthermore, we investigated in great detail the basic building block of the simulator and demonstrated that the correlated hopping for the single particle case can be realised efficiently when the resonance condition $\vert \Delta \vert = \hbar \Omega_R$ is fulfilled. Interestingly, we have found that the impurity trap plays only a marginal role for the correlated hopping process and that for atom-ion systems the scheme is resilient to micro-motion. Furthermore, we have shown that in the limit of a large number of bosons the system can be used to explore the physics of the Higgs model. In the future it would be interesting to investigate more closely the impurity-induced correlated many-body quantum dynamics and/or when the initial superfluid state in the Bose-Hubbard is suddenly perturbed by the presence of the impurities via state-dependent hopping terms as well as gauge plaquette terms that appear in higher spatial dimensions. 


\section*{Acknowledgements} 

We gratefully acknowledge A. Alberti, M. Dalmonte, R. Gerritsma, S. Montangero, and M. Valiente for valuable feedback on the manuscript, A. B. Michelsen for performing the optimal control calculation, and J. M. Schurer for discussions (AN). This work was supported by The Hamburg Centre for Ultrafast Imaging (AN), the UPV/EHU grant EHUA15/17, the Spanish Ministerio de Econom\'ia y Competitividad through the project MINECO/FEDER FIS2015-69983-P, Basque Government IT986-16 (ER), and the Danish Council for Independent Research under the Sapere Aude program (ASD and NTZ).

\appendix


\section{Analytical solutions to the Double Harmonic Potential}

\begin{figure*}
\includegraphics[width=1.0\textwidth]{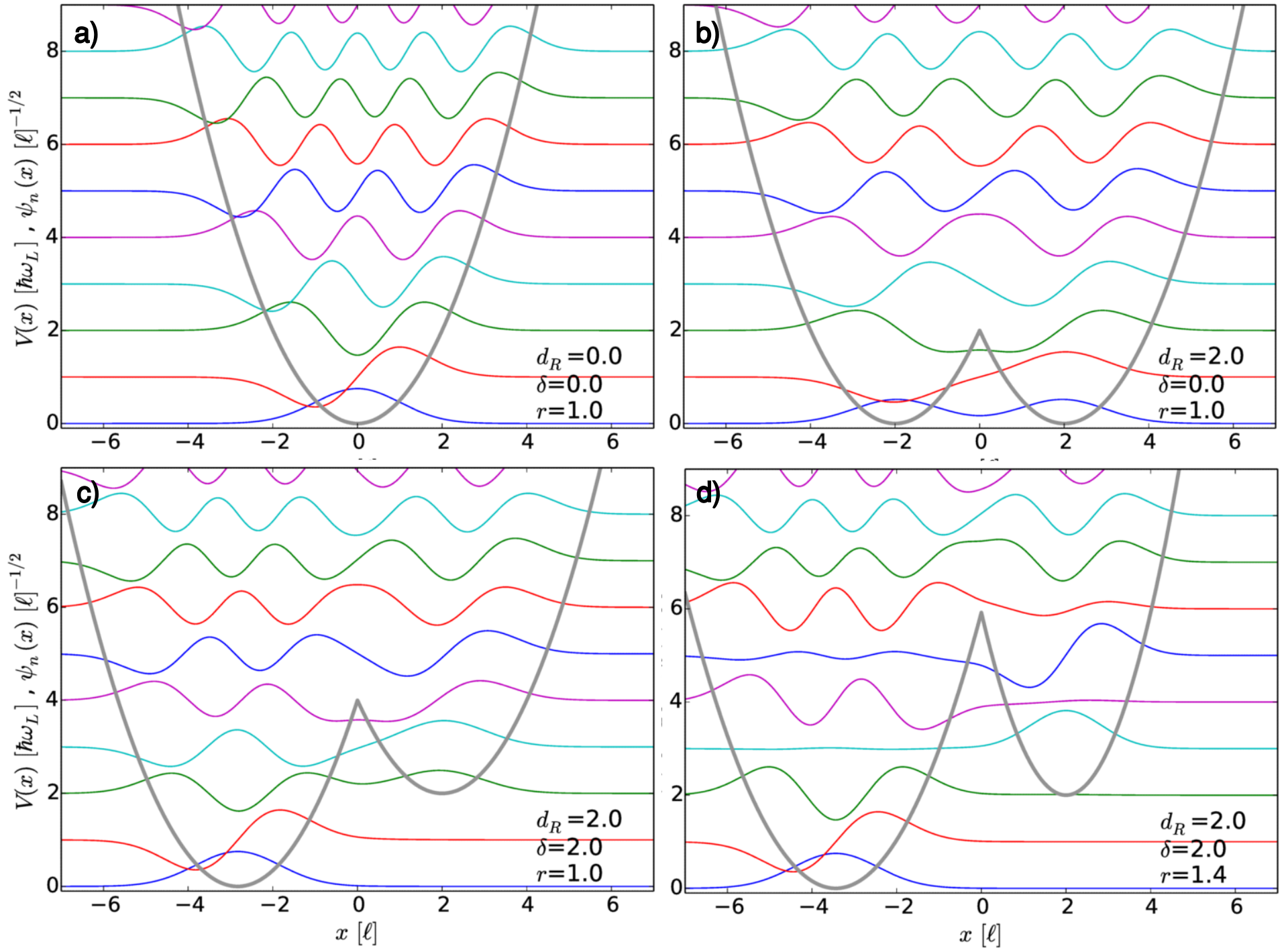}
\caption{a) Eigenfunctions of the harmonic potential, $\psi_n(x)$, for $d_R=0$, $\delta=0$, and $r=1$. b) Eigenfunctions for the symmetric double harmonic potential $d_R=2$, $\delta=0$, and $r=1$. c) Eigenfunctions for the tilted double harmonic potential for $d_R=2$, $\delta=2$, and $r=1$. d) Eigenfunctions for a narrower tilted double harmonic potential, namely for $d_R=2$, $\delta=2$, and $r=1.4$. For the sake of better visibility, all eigenfunctions have been located on the vertical (energy) axis separated by one $\hbar\omega_L$. Besides, in all panels $d_R$ and $\delta$ are in units of $\ell$ and $\hbar\omega_L$, respectively.}
\label{fig1}
\end{figure*}

We consider a particle in a double harmonic potential as illustrated in Fig.~\ref{fig1} (thick black line), whose mathematical expression is given by:

\begin{equation}
V_{\text p}^{\mathrm{ext}}(x)= 
\begin{cases}
\frac{1}{2}m_{\text p}\omega_L^2(x+d_L)^2,& \text{if } x<0\\
\frac{1}{2}m_{\text p}\omega_R^2(-x+d_R)^2+\Delta,& \text{otherwise}
\end{cases}
\end{equation}
where $d_L$ and $d_R$ are the centres of the double harmonic potential one in left and the other in right of $x=0$, respectively. Here $\omega_L$ and $\omega_R$ are the trapping frequencies, $m_{\text{p}}$ is the mass of the trapped particle, and $\Delta$ is the height difference between the two harmonic potentials (i.e., the energy offset). Hereafter we shall assume that the potential is a continuous function. Thus, we must have $\frac{1}{2}m_{\text p}\omega_L^2d_L^2=\frac{1}{2}m_{\text p}\omega_R^2d_R^2+\Delta$.

Our goal is to solve the following eigenvalue problem 

\begin{equation}
\label{eq:SE}
-\frac{\hbar^2}{2m_{\text p}}\frac{\partial^2\psi(x)}{\partial x^2}+V_{\text p}^{\mathrm{ext}}(x)\psi(x)=E\psi(x)
\end{equation}
For the sake of convenience, we introduce the length $\ell=\sqrt{\hbar/(m_{\text p}\omega_L)}$ and energy $\epsilon=\hbar\omega_L$ scales. By transforming all the lengths with respect to $\ell$ and the energies with respect to $\epsilon$, and by rewriting the energy offset as $\Delta\equiv\hbar\omega_L\delta$, the continuity condition becomes $d_L^2=r^2\cdot d_R^2+2\delta$ where $r\equiv\omega_R/\omega_L$. In the rescaled units the Schr\"odinger equation for $x<0$ reads:

\begin{equation}
\frac{\partial^2\psi(x)}{\partial x^2}-\left[(x+d_L)^2-2E\right]\psi(x)=0.
\end{equation}
By defining $E=\nu_1+\frac{1}{2}=r\cdot(\nu_2+\frac{1}{2})+\delta$, where $\nu_1$ and $\nu_2$ are real numbers (not necessary integers) to be determined, the above equation turns into:

\begin{equation}
\frac{\partial^2\psi(x)}{\partial x^2}-\left[\left(x+d_R\right)^2-2\left(\nu_1+\frac{1}{2}\right)\right]\psi(x)=0.
\end{equation}
Hereafter, we introduce the new variable $y = -\sqrt{2}(x+d_L)$, which turns the above equation into:

\begin{equation}
\frac{\partial^2\psi(y)}{\partial y^2}-\left[\frac{y^2}{4}-\left(\nu_1+\frac{1}{2}\right)\right]\psi(y)=0.
\label{transformedeq}
\end{equation}
However, note that the introduced variable, $y$, could also be defined as $y = \sqrt{2}(x+d_L)$, which is a solution to the differential equation, too. Nevertheless, only $y = -\sqrt{2}(x+d_L)$ gives a normalised and convergent wave function within the interval $(-\infty,0)$, whereas the former does not. We note that Eq.~(\ref{transformedeq}) has the form of a parabolic cylinder differential equation

\begin{equation}
\frac{\partial^2w}{\partial z^2}-\left(\frac{z^2}{4}+a\right)w=0,
\end{equation}
whose solutions are named parabolic cylinder functions and denoted as $D_{-a-\frac{1}{2}}(z)$~\cite{Abramowitz:1992}. In order to solve our eigenvalue problem, we shall make use of the following property of such functions

\begin{align}
\frac{\partial D_{\nu}}{\partial z}&=\frac{1}{2}\nu D_{\nu-1}(z)-\frac{1}{2}D_{\nu+1}(z),
\end{align}
which enable us to obtain the following solutions for the differential equation~(\ref{transformedeq}):

\begin{equation}
\psi(x)= 
\begin{cases}
c_1 D_{\nu_1}\left(-\sqrt{2}(x+d_L)\right),& \text{if } x<0\\
c_2 D_{\nu_2}\left(\sqrt{2r}(x-d_R)\right) & \text{otherwise}.
\end{cases}
\end{equation}
Here $c_1$ and $c_2$ are some constants, to be determined later in order to normalise the wave function. The derivative of such wave function yields
\begin{equation*}
\frac{\partial\psi(x)}{\partial x}= 
\begin{cases}
\frac{-1}{\sqrt{2}} c_1 \left\{\nu_1D_{\nu_1-1}(z_1)-D_{\nu_1+1}(z_1)\right\},& \text{if } x<0\\
\sqrt{\frac{r}{2}} c_2 \left\{\nu_2D_{\nu_2-1}(z_2)-D_{\nu_2+1}(z_2)\right\},& \text{otherwise}
\end{cases}
\end{equation*}
where $z_1\equiv-\sqrt{2}(x+d_L)$ and $z_2\equiv\sqrt{2r}(x-d_R)$. In order to find the unknowns, $c_1$, $c_2$, $\nu_1$ and $\nu_2$, one must impose the following conditions:

\begin{enumerate}
\item $\psi(x)|_{x=0}$ must be continuous.
\item $\psi(x)'|_{x=0}$ must be continuous.
\item Energy is the same: $\nu_1+1/2=r(\nu_2+1/2)+\delta$.
\item Normalisation: $1=\int |\psi(x)|^2 dx$.
\end{enumerate}
If the potentials are symmetric around $x=0$, one can use the fact that the even solutions must have $\psi(x)'|_{x=0}=0$ and the odd solutions $\psi(x)|_{x=0}=0$. In general, however, one must derive additional (transcendental) equations in order to determine the unknowns. Special solutions are exemplarily displayed in Fig.~\ref{fig1}.

When the particle in the double well interacts with the (static) impurity located in $x=0$, we have to incorporate in the Scr\"odinger equation~(\ref{eq:SE}) the following additional (external) potential 
\begin{align}
\label{eq:Vx}
V_s(x) = \sum_{s=\uparrow,\,\downarrow}\left\{ \upsilon_{\mathrm{1D}}^{e,s}(x - x_{\text i}) + \upsilon_{\mathrm{1D}}^{o,s}(x - x_{\text i})\right\},
\end{align}
where 
\begin{align}
\label{eq:pseudo-eo}
\upsilon_{\mathrm{1D}}^{e,s}(x) = g_{\mathrm{1D}}^{e,s} \hat\delta_{\pm}(x),\qquad
\upsilon_{\mathrm{1D}}^{o,s}(x) = g_{\mathrm{1D}}^{o,s} \delta^{\prime}(x)\hat\partial_{\pm}.
\end{align}
Here $g_{\mathrm{1D}}^{e,s} = -\hbar^2/\mu a_{\mathrm{1D}}^{e,s}$ and $g_{\mathrm{1D}}^{o,s} = -\hbar^2 a_{\mathrm{1D}}^{o,s}/\mu$ with $\mu$ being the reduced particle-impurity mass, and $a_{\mathrm{1D}}^{e,s},\,a_{\mathrm{1D}}^{o,s}$ are the 1D spin-dependent scattering lengths for even- and odd-waves, respectively. Besides, the action of the two operators appearing in Eq.~(\ref{eq:pseudo-eo}) on a wave-function $\psi(x)$ is given by:

\begin{align}
2\, \hat\delta_{\pm}(x)\psi(x) &= [\psi(0^+) + \psi(0^-)] \delta(x),\nonumber\\
2\, \hat\partial_{\pm}\psi(x) &= [\psi^{\prime}(0^+) + \psi^{\prime}(0^-)] ,
\end{align}
where $\psi(0^{\pm}) = \lim_{x\rightarrow 0^{\pm}}\psi(x)$. With these conditions we end up with:

\begin{align}
[\psi^{\prime}(0^+) - \psi^{\prime}(0^-)] &= -\frac{1}{a_{1D}^e}[\psi(0^+) + \psi(0^-)],\\
[\psi(0^+) - \psi(0^-)] &= -a_{1D}^o [\psi^{\prime}(0^+) + \psi^{\prime}(0^-)] ,
\end{align}
for even and odd scattering, respectively. 

By using these conditions instead of the above outlined conditions 1. and 2. of the non-interacting particle problem, we end up with the following transcendental equation:
\begin{align}
&\frac{\frac{1}{a_{\text{1D}}^e}D_{\nu_1}(z_1)+\sqrt{2}\left. \frac{\partial D_{\nu_1}}{\partial z}\right|_{z_1}}{D_{\nu_1}(z_1)+a_{\text{1D}}^o \sqrt{2}\left. \frac{\partial D_{\nu_1}}{\partial z}\right|_{z_1}}+\frac{\frac{1}{a_{\text{1D}}^e}D_{\nu_2}(z_2)+{\sqrt{2r}}\left. \frac{\partial D_{\nu_2}}{\partial z}\right|_{z_2}}{D_{\nu_2}(z_2)+a_{\text{1D}}^o \sqrt{2r}\left. \frac{\partial D_{\nu_2}}{\partial z}\right|_{z_2}}\nonumber\\
&=0, 
\label{transcendental-eq}
\end{align}
where $z_1=-\sqrt{2}d_L$ and $z_2=-\sqrt{2r}d_R$, and from condition 3. we arrive at $\nu_1=r(\nu_2+1/2)+\delta-1/2$. Hence, we are basically left with one equation with one unknown, $\nu_2$. Therefore we can define the left-hand-side of Eq.~(\ref{transcendental-eq}) as being a function of $\nu_2$, $f(\nu_2)$, which we have to solve as $f(\nu_2)=0$ as a final step. This is done by finding the zeros of such a function numerically. Once the sequence of real numbers $\nu_2$ are found, the corresponding $\nu_1$ are obtained accordingly. Finally, by using the normalisation condition 4., we can obtain the constants $c_1$ and $c_2$.\\


\subsection{Limiting cases}

Let us note that in the case for which $r=1$, $\delta=0$, $d_L=d_R=0$, we have $\nu_1=\nu_2=\nu$. Moreover, if $a_{\text{1D}}^o=0$ and $-1/a_{\text{1D}}^e\equiv g^e_{\text{1D}}$, we obtain the non-trivial result calculated for two interacting atoms in a harmonic 
trap~\cite{Busch:1998}: 
\begin{align}
\frac{-1}{g^e_{\text{1D}}}=\frac{1}{2}\frac{\Gamma{(-E/2+1/4)}}{\Gamma{(-E/2+3/4)}}.
\end{align}\\
In the other limit, that is when $a_{\text{1D}}^o\equiv -g^o_{\text{1D}}$ and $1/a_{\text{1D}}^e=0$, we have
\begin{align}
\frac{-1}{g^o_{\text{1D}}}=2\frac{\Gamma{(-E/2+3/4)}}{\Gamma{(-E/2+1/4)}}
\end{align}
Note that in both limits $E=\nu+1/2$ holds.


\section{Bosonic Josephson junction in the large particle limit}

In this section we would like to show in a more rigorous way that, within the two-mode approximation, when the particle number per well is sufficiently large, the ensemble of bosons can be treated as a Bose-Einstein condensate, and therefore the Hamiltonian term

\begin{equation}
\hat H_{\text{2m}} = J_z \hat b^{\dagger}_{L}  \tilde \sigma^{+} \hat b_{R}  + \text{h.c.}  
\end{equation}
can be replaced by Eq.~(\ref{eq:BJJ}). To this end, let us consider $N$ particles in both wells with a well-defined phase difference $\theta= \theta_{L} - \theta_{R}$, with $N \gg \Delta N$, i.e., the total number of particles is much greater than any variation due to the dynamics. In this scenario, the initial state of the compound bosonic ensemble and impurity spin system is given by
\begin{equation}
\begin{split}
&|N,\theta\rangle \otimes |\sigma \rangle = \frac{1}{\sqrt{N!}} \frac{1}{2^{N/2}}  \left( e^{i \theta_{L}} \hat b^{\dagger}_{L} + e^{i \theta_{R}} \hat b^{\dagger}_{R}  \right)^{N} |\text{vac} \rangle  \otimes |\sigma \rangle\\
&= \sum_{n=0}^{N} \frac{ e^{i N \theta_{R}} \sqrt{N!} e^{i n \left( \theta_{L} -  \theta_{R} \right) } }{n! \left(N-n\right) ! 2^{N/2}}   \left(\hat b^{\dagger}_{L} \right)^{n}  \left(\hat b^{\dagger}_{R}\right)^{N-n}   |\text{vac} \rangle  \otimes |\sigma \rangle,
\end{split}
\end{equation}
where $\ket{N,\theta}$ denotes the state of the bosonic ensemble with $N$ particles and a given phase difference $\theta$, $\ket{\text{vac}}$ the vacuum state, and $\ket{\sigma}$ the spin state of the impurity. The distribution $\frac{1}{n! \left(N-n\right) !}$ is exponentially picked at $n=N/2$. When $N \gg 1$, the state characterises two condensates of $\frac{N}{2}$ particles in each well, i.e., 
\begin{align*}
|N,\theta\rangle \sim e^{i N \theta_{L} /2} \left(b^{\dagger}_{L} \right)^{N/2} e^{-i N \theta_{R} /2} \left(b^{\dagger}_{R} \right)^{N/2} |\text{vac}\rangle.
\end{align*}
By exploiting the identity $[ \hat b , ( \hat b^{\dagger})^{n} ] = n ( \hat b^{\dagger} )^{n-1}$, the action of the Hamiltonian $\hat H_{\text{2m}}$ on this state is given by
\begin{widetext} 
\begin{equation}
\begin{split}
&\hat H_{\text{2m}} |N,\theta\rangle \otimes |\sigma \rangle=\left( \hat b^{\dagger}_{L} \tilde \sigma^{+} \hat b_{R} + \hat b^{\dagger}_{R} \tilde \sigma^{-} \hat b_{L} \right) |N,\theta\rangle \otimes |\sigma \rangle \\
&= J_z\frac{ \sqrt{N!} e^{i N \theta_{R}} }{2^{N/2}} \sum_{n} \left[ \frac{e^{i n \left( \theta_{L} -  \theta_{R} \right) }}{n! \left(N-n-1\right) !}   \left(\hat b^{\dagger}_{L} \right)^{n+1} \tilde \sigma^{+} \left(\hat b^{\dagger}_{R}\right)^{N-n-1} + \frac{e^{i n \left( \theta_{L} -  \theta_{R} \right) }}{\left(n-1\right)! \left(N-n\right) !}   \left(\hat b^{\dagger}_{L} \right)^{n-1} \tilde \sigma^{-}  \left(\hat b^{\dagger}_{R}\right)^{N-n+1}  \right] |\text{vac} \rangle  \otimes |\sigma \rangle\\
&= J_z \frac{ \sqrt{N!} e^{i N \theta_{R}} }{2^{N/2}} \sum_{m} \left[ \frac{m e^{i \left( m-1\right) \left( \theta_{L} -  \theta_{R} \right) }}{ m ! \left(N-m\right) !}   \left(\hat b^{\dagger}_{L} \right)^{m}  \tilde \sigma^{+} \left(\hat b^{\dagger}_{R}\right)^{N-m} + \frac{\left(N-m \right) e^{i \left( m+1\right) \left( \theta_{L} -  \theta_{R} \right) }}{ m! \left(N-m\right) !}   \left(\hat b^{\dagger}_{L} \right)^{m} \tilde \sigma^{-} \left(\hat b^{\dagger}_{R}\right)^{N-m}  \right] |\text{vac} \rangle  \otimes |\sigma \rangle \\
& \sim J_z\frac{N}{2} \left( \tilde \sigma^{+} e^{- i \left( \theta_{L} -  \theta_{R} \right) } + \tilde \sigma^{-} e^{i \left( \theta_{L} -  \theta_{R} \right) } \right) |N,\theta\rangle \otimes |\sigma \rangle.
\end{split}
\end{equation}
\end{widetext} 
Hence, this shows that the state $|N,\theta\rangle \otimes |\sigma \rangle$ is almost an eigenstate of the Hamiltonian $\hat H_{\text{2m}}$. In particular, because of the last equality in the equation above, the operators $b^{\dagger}_{L,R}$ become proportional to $\sqrt{N} e^{i\theta_{L,R}}$ in the large-$N$ limit, thus demonstrating that in such a case the effective Hamiltonian is given by Eq.~(\ref{eq:BJJ}).


\section{Effective equation of motion for a condensate coupled to a spin impurity}

In this section we derive the equations of motion for a bosonic ensemble in interaction with a quantum spin, thus without motional degrees of freedom for the latter. In particular, we focus on the simplest scenario for which the bosonic ensemble is in the condensate state (i.e., mean field theory), but we take into account all correlations between the ensemble and the spin system.


\subsection{Many-body Hamiltonian and effective equations of motion}

The Hamiltonian describing $N$ interaction bosons of mass $m$ with a spin-$\frac{1}{2}$ impurity is given by

\begin{align}
\label{eq:H}
\hat H & = \sum_{j=1}^N\left[ \frac{\hat p_j^2}{2 m}+V_{\text{ext}}(x_j)+\sum_{\alpha=0}^1U_{\alpha}(x_j)\ket{\alpha}\bra{\alpha}\right]\nonumber\\
\phantom{=}&+\frac{g}{2}\sum_{i\ne j}\delta(x_i-x_j)+\frac{\hbar\Omega_R}{2}\hat\sigma_x
\end{align}
with $\hat p_j =- i\hbar\frac{\partial}{\partial x_j}$ being the momentum operator of the $j$-th boson, $V_{\text{ext}}(x_j)$ the external potential (e.g., the double well), $U_{\alpha}(x_j)$ a potential that depends on the internal state $\ket{\alpha}$ of the impurity (in our setting the contact potentials), $\Omega_R$ the Rabi frequency, and $\hat\sigma_x = \ket{0}\bra{1} + \ket{1}\bra{0}$ the Pauli matrix. The coupling constant $g$ describes the interaction among the bosons in quasi-1D, which typically depends on the three-dimensional s-wave scattering length and the transverse trap frequency. Note that if $U_{\alpha}(x_j)=0$ $\forall\alpha$, then no coupling between the spin impurity and the ensemble of bosons exists. This implies that the two subsystems are fully uncorrelated and that the full state is simply given by the tensor product of a state living in the Hilbert space associated to the ensemble of bosons and a state of the impurity in $\mathbb{C}^2$.


\subsection{Many-body state ansatz}

The most general state for such an ensemble of $N$ bosons coupled to a two-level quantum system is given by 

\begin{align}
\ket{\Psi} = \sum_{\alpha=0}^1 c_{\alpha}\ket{\Phi_{\alpha}}\ket{\alpha}
\end{align}
with $\vert c_0\vert^2 + \vert c_1\vert^2 =1$. Such a state encompasses correlations among the bosons as well as between the bosonic ensemble and the spin impurity. Next, we assume that the $N$-particle state is a product state for each spin component (mean-field approximation), namely $\ket{\Phi_{\alpha}} = \prod_{j=1}^N\ket{\varphi_{\alpha}^{(j)}}$. Thus, the above outlined ansatz reduces to 

\begin{align}
\label{eq:mbs-mf}
\ket{\Psi} = \sum_{\alpha=0}^1 c_{\alpha} \prod_{j=1}^N\ket{\varphi_{\alpha}^{(j)}}\ket{\alpha}
\end{align}
with $\braket{\varphi_\alpha}{\varphi_\alpha}=1$ $\forall\alpha$.


\subsection{Condensate-impurity equations of motion:}

We aim at determining the equations of motion for the coefficients $c_{\alpha}$ as well as for the condensate wave functions $\varphi_{\alpha}(x)$. To this end, we employ the so-called Dirac-Frankel variational principle, $\bra{\delta\Psi}(i\hbar\partial_t - \hat H)\ket{\Psi}=0$, where $\delta\Psi$ denotes the variation of the many-body wave function with respect to the free parameters $c_{\alpha}$ and $\varphi_{\alpha}$. 

After a lengthy calculation we arrive at the equations of motion for the spin-impurity coefficients ($\overline{\alpha} = 1- \alpha$)

\begin{align}
\label{eq:dc}
&i\hbar\left[
\dot c_{\alpha} + N c_{\alpha} \braket{\varphi_\alpha}{\dot\varphi_\alpha}
\right] = N c_\alpha \braket{\varphi_\alpha}{\hat H_0^\alpha\vert\varphi_\alpha}\nonumber\\
\phantom{} &
+\frac{g}{2} N (N-1) c_\alpha \braket{\varphi_\alpha,\varphi_\alpha}{\delta(x- y)\vert\varphi_\alpha,\varphi_\alpha}\nonumber\\
\phantom{}&
+c_{\overline{\alpha}}\hbar\Omega_R\left(
\braket{\varphi_\alpha}{\varphi_{\overline{\alpha}}}
\right)^N
\end{align}
with $\hat H_0^\alpha = \frac{\hat p^2}{2m}+V_{\text{ext}}+U_{\alpha}$, whereas for the condensate wave functions we obtain

\begin{widetext}
\begin{align}
\label{eq:dphi}
&i\hbar\left\{
c_{\alpha}^* \dot c_{\alpha} \ket{\varphi_\alpha}+\vert c_{\alpha}\vert^2\left[
\ket{\dot\varphi_\alpha} + (N-1) \braket{\varphi_\alpha}{\dot\varphi_\alpha}\ket{\varphi_\alpha}
\right]
\right\}= \vert c_\alpha\vert^2\left\{
\hat H_0^\alpha+
(N-1)\braket{\varphi_\alpha}{\hat H_0^\alpha\vert\varphi_\alpha}
+\frac{g}{2}\left[
2(N-1)\braket{\varphi_\alpha}{\delta(x-y)\vert\varphi_\alpha}\right.\right.\nonumber\\
&\phantom{}
\left.\left.+(N^2-3N+2)  \braket{\varphi_\alpha,\varphi_\alpha}{\delta(x- y)\vert\varphi_\alpha,\varphi_\alpha}
\right]
\right\}\ket{\varphi_\alpha}
+c_\alpha^*c_{\overline{\alpha}}\hbar\Omega_R\left(\braket{\varphi_\alpha}{\varphi_{\overline{\alpha}}}\right)^{N-1} \ket{\varphi_{\overline{\alpha}}}
\end{align}
\end{widetext}
with $\braket{\varphi_\alpha}{\delta(x-y)\vert\varphi_\alpha} = \vert\varphi_\alpha(x)\vert^2$, and

\begin{align}
\braket{\varphi_\alpha,\varphi_\alpha}{\delta(x- y)\vert\varphi_\alpha,\varphi_\alpha}  = \int_{\mathbb{R}}d x \vert\varphi_\alpha(x)\vert^4.
\end{align}
Now, by substituting $\dot c_{\alpha}$  in Eq.~(\ref{eq:dphi}) with Eq.~(\ref{eq:dc}) we finally arrive at 

\begin{align}
\label{eq:dphi2}
i\hbar\ket{\dot\varphi_\alpha}  &= +i\hbar \braket{\varphi_\alpha}{\dot\varphi_\alpha}\ket{\varphi_\alpha}+\hat H_{\text{gp}}^\alpha[\varphi_\alpha]\ket{\varphi_\alpha} \nonumber\\
\phantom{=} & + \frac{c_\alpha^*c_{\overline{\alpha}}}{\vert c_\alpha\vert^2}\hbar\Omega_R\braket{\varphi_\alpha}{\varphi_{\overline{\alpha}}}^{N-1} \ket{\varphi_{\overline{\alpha}}}\nonumber\\
\phantom{=} & - \left( 
\langle \hat H_{\text{gp}}^\alpha[\varphi_\alpha] \rangle + 
\frac{c_\alpha^*c_{\overline{\alpha}}}{\vert c_\alpha\vert^2}\hbar\Omega_R \braket{\varphi_\alpha}{\varphi_{\overline{\alpha}}}^{N} 
\right)\ket{\varphi_\alpha},
\end{align}
where $\hat H_{\text{gp}}^\alpha[\varphi_\alpha] = \frac{\hat p^2}{2m} + V_{\text{ext}}+U_{\alpha}+ g (N-1) \vert \varphi_{\alpha}\vert^2$ and 
$\langle \hat H_{\text{gp}}^\alpha[\varphi_\alpha] \rangle = \braket{\varphi_\alpha}{\hat H_{\text{gp}}[\varphi_\alpha]\vert \varphi_\alpha}$. With these definitions, we can rewrite Eq.~(\ref{eq:dc}) as

\begin{align}
\label{eq:dc2}
i\hbar
\dot c_{\alpha} & = c_{\overline{\alpha}}\hbar\Omega_R
\braket{\varphi_\alpha}{\varphi_{\overline{\alpha}}}^N \nonumber\\
\phantom{=} &
+N c_\alpha \left(\langle\hat H_{\text{gp}}^\alpha[\varphi_\alpha/\sqrt{2}]\rangle
 - i\hbar  \braket{\varphi_\alpha}{\dot\varphi_\alpha}
 \right).
\end{align}

To further simplify the above outlined equations of motion, we first perform a unitary transformation on the coefficients, that is, $\mathbf{c} = \hat U \mathbf{C}$ with $\mathbf{c}\equiv (c_0,c_1)^T$ (similarly for $\mathbf{C}$), where 

\begin{align}
\hat U = \left(
\begin{array}{cc}
e^{-i\eta_0(t)} & 0 \\
0 & e^{-i\eta_1(t)}
\end{array}
\right). 
\end{align}
By applying this unitary, we get

\begin{align}
\label{eq:dC2}
i
\dot C_{\alpha} & = C_{\overline{\alpha}} 
e^{-i[\eta_{\overline{\alpha}}(t) - \eta_\alpha(t)]}
\Omega_R
\braket{\varphi_\alpha}{\varphi_{\overline{\alpha}}}^N, \\
\eta_\alpha(t) &= \frac{N}{\hbar} \int_0^td\tau  \left(\langle\hat H_{\text{gp}}^\alpha[\varphi_\alpha/\sqrt{2}]\rangle
 - i\hbar  \braket{\varphi_\alpha}{\dot\varphi_\alpha}
 \right) \nonumber\\
 \phantom{=}&= \frac{N}{\hbar} f_\alpha(t).
\end{align}
Now, let us note that $\langle\hat H_{\text{gp}}^\alpha[\varphi_\alpha/\sqrt{2}]\rangle$ is a real number, as it corresponds to the expectation value of a (nonlinear) hermitian operator, but that also $i\hbar  \braket{\varphi_\alpha}{\dot\varphi_\alpha}$ is a real number. To see this, let us note that 
$\frac{d}{dt} \langle \varphi_\alpha \vert \varphi_\alpha \rangle = 2\Re\{ \braket{\varphi_\alpha}{\dot\varphi_\alpha}\} = 0$. This implies that $\braket{\varphi_\alpha}{\dot\varphi_\alpha}$ is purely imaginary and as a consequence $i\hbar  \braket{\varphi_\alpha}{\dot\varphi_\alpha}$ is purely real. 
Hence, $\eta_\alpha(t)$ is a good phase definition. Thus, by defining $\ket{\phi_\alpha}:=\ket{\varphi_\alpha e^{-\frac{i}{\hbar} f_\alpha(t)}}$, we have

\begin{align}
\label{eq:dcfinal}
i\dot C_{\alpha} & = C_{\overline{\alpha}} 
\Omega_R
\braket{\phi_\alpha}{\phi_{\overline{\alpha}}}^N.
\end{align}
Now, since $\ket{\varphi_\alpha}= e^{\frac{i}{\hbar} f_\alpha(t)}\ket{\phi_\alpha}$, we can use Eq.~(\ref{eq:dphi2}) in order to get a differential equation for $\ket{\phi_\alpha}$. After a straightforward calculation, one arrives at

\begin{align}
\label{eq:dphifinal}
i\hbar\ket{\dot\phi_\alpha}  &= \hat H_{\text{gp}}^\alpha[\phi_\alpha]\ket{\phi_\alpha} + \frac{C_\alpha^*C_{\overline{\alpha}}}{\vert C_\alpha\vert^2}\hbar\Omega_R\braket{\phi_\alpha}{\phi_{\overline{\alpha}}}^{N-1} \ket{\phi_{\overline{\alpha}}}\nonumber\\
\phantom{=} & - \left( 
\frac{g}{2}(N-1)\langle \vert\phi_\alpha\vert^2\rangle + 
\frac{C_\alpha^*C_{\overline{\alpha}}}{\vert C_\alpha\vert^2}\hbar\Omega_R \braket{\phi_\alpha}{\phi_{\overline{\alpha}}}^{N} 
\right)\ket{\phi_\alpha}.
\end{align}
Alternatively, we can combine Eq.~(\ref{eq:dcfinal}) and Eq.~(\ref{eq:dphifinal}) in a single one by defining $\ket{\psi_\alpha} := C_\alpha \ket{\phi_\alpha}$. This yields

\begin{align}
\label{eq:dpsiNum}
i\hbar\ket{\dot\psi_\alpha}  &= \hat H_{\text{gp}}^\alpha[\phi_\alpha]\ket{\psi_\alpha} +\hbar\Omega_R\braket{\phi_\alpha}{\phi_{\overline{\alpha}}}^{N-1} \ket{\psi_{\overline{\alpha}}}\nonumber\\
\phantom{=} & - 
\frac{g}{2}(N-1)\langle \vert\phi_\alpha\vert^2\rangle \ket{\psi_\alpha}
\end{align}
with $\ket{\phi_\alpha} = \ket{\psi_\alpha}/\sqrt{\braket{\psi_\alpha}{\psi_\alpha}}$. In this way, one can simulate a single system of coupled differential equations~(\ref{eq:dpsiNum}), and therefore forget about the equation of motion of the coefficients $C_\alpha$. Note that $\sum_{\alpha = 0}^1\int_{\mathbb{R}} dx \vert \psi_\alpha(x,t)\vert^2 = 1$, and thereby $\int_{\mathbb{R}} dx \vert \psi_\alpha(x,t)\vert^2\ne 1$.


\section{Correlated hopping in the many-body mean-field scenario}

\begin{figure}[t]
\includegraphics[width=\linewidth]{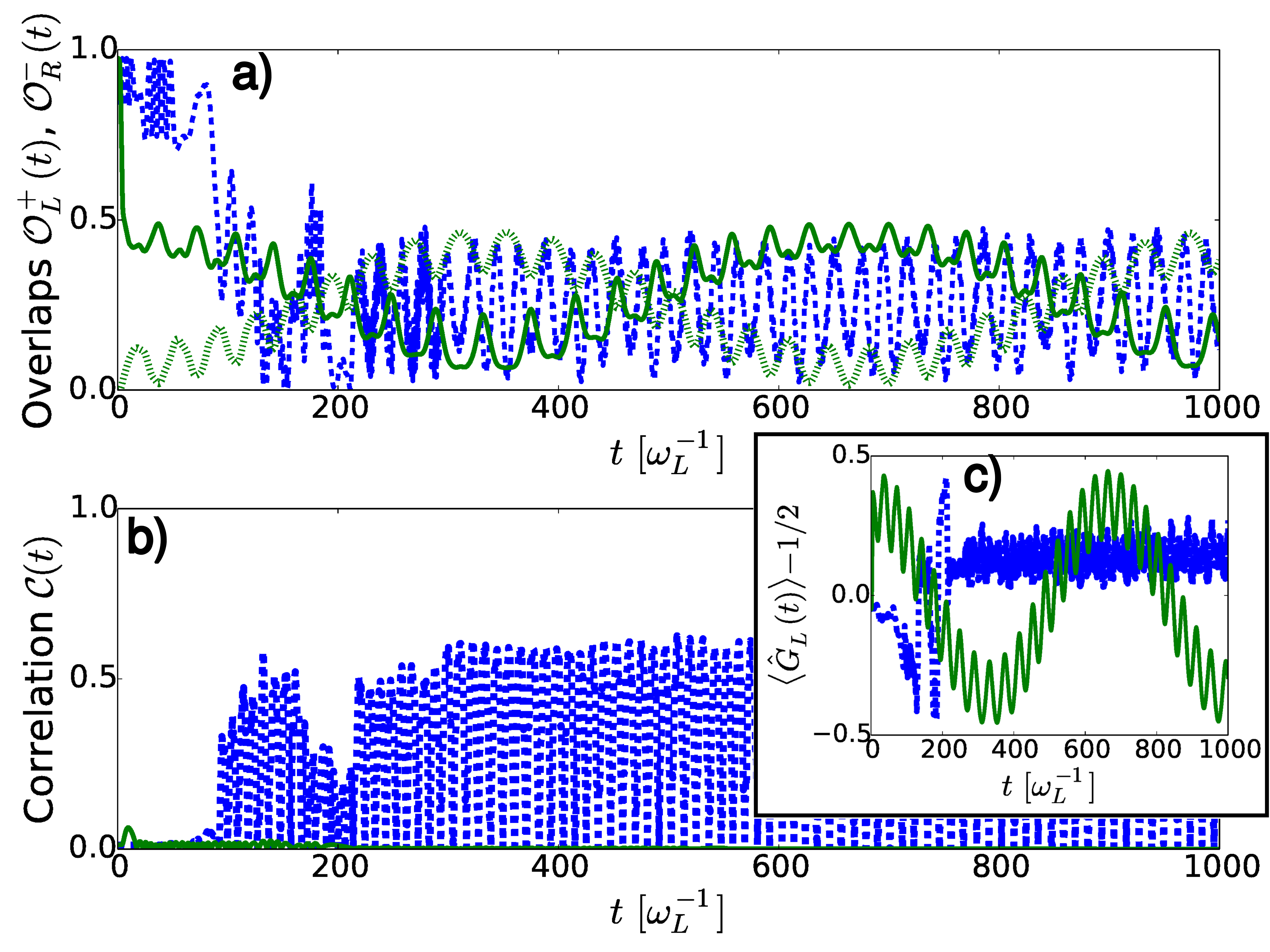}
\caption{(Colour online). Green solid ($\mathcal{O}_L(t)$) and dotted ($\mathcal{O}_R(t)$) lines show the result of the correlated hopping with a condensate by means of the compensating control pulse technique~(\ref{eq:Rabi}), while blue dashed line shows the results from machine learning techniques. The simulation corresponds to $N=10$, $^{87}$Rb bosonic atoms with an interaction strength $g=0.21 \hbar\omega_L\ell$ that assumes $\omega_L = 2 \pi~1$ kHz and a transverse trap frequency $10 \omega_L$. Panel a) shows the overlap fidelity, panel b) the correlation function and panel c) the expectation value of the operator $\hat G_L$.}
\label{supp:fig5}
\end{figure}

We have investigated numerically the possibility to realise the correlated hopping, as for the single particle case discussed in Sec. \ref{sim-neutral}, when the system is an ensemble of ultra-cold bosons described in the mean-field approximation, as discussed in the previous appendix. To this end, we solved the equations of motion~(\ref{eq:dpsiNum}). In Ref.~\cite{Brouzos:2015} it has been shown that for a bosonic Josephson junction without contact to a spin impurity, whereby the condensate is prepared initially in one of the two wells, the compensating control pulse technique allows to transfer the entire condensate to the other well. Such a pulse, that represents the energy offset between the two wells, has a simple analytical form: $\Delta(t) = - \Delta U - U \cos(J t)$. Here $J = (E_R^0 - E_L^0)/\hbar$, $\Delta U = U_L - U_R$, $U = (U_L + U_R)/2$,

\begin{align}
U_{L} = g \int_{\mathbb{R}}dx\vert\langle x\vert L\rangle\vert^4,
\end{align}
and similarly for $U_R$. Inspired by this work, we modulated the Rabi frequency by using the following compensating control pulse:

\begin{align}
\label{eq:Rabi}
\hbar\Omega_R(t) = E_L^0 - E_R^0 + \Delta U + U  \cos(J t).
\end{align}
Contrarily to the case without spin impurity, we need to modulate the Rabi frequency over time in order to fulfil the resonance condition that, because of the inter-particle interaction, changes in time as the condensate tunnels through the barrier. The result of such a transfer process is illustrated in Fig.~\ref{supp:fig5}. As it is shown, the condensate transfer does not work very well (roughly 50\% only) and, importantly, the correlated hopping is not occurring at all, as other (undesired) states are strongly populated. 

We then used the pulse~(\ref{eq:Rabi}) as initial guess for the Rabi frequency for an optimal control study and at the same time we allowed control for the energy offset $\Delta$. The outcome of such an investigation is illustrated in Fig.~\ref{supp:fig5}. As it can be seen, even by varying two experimentally accessible parameters via optimal control theory we were unable to realise the desired process. Even though this is not a rigorous mathematical proof, it strongly indicates that the correlated hopping in presence of interactions is substantially modified. Moreover, the process seems not attainable within a mean-field description. This might be also due to the fact that our derivation of the target gauge invariant Hamiltonian relies on the single particle tunnelling and that in the many-body scenario high-order tunnelling effects might play a crucial role, while in our description such effects are not allowed. Indeed, as it has been observed for a purely bosonic Josephson junction without spin impurity, at times close to the so-called quantum speed limit the mean-field description breaks down~\cite{Brouzos:2015}. Hence, it would be interesting to understand, whether in the presence of a gauge field, the process can take place by means of a concerted action of inter- and intra-particle correlations, a task that we shall pursue in the future.


\begin{thebibliography}{44}
\expandafter\ifx\csname natexlab\endcsname\relax\def\natexlab#1{#1}\fi
\expandafter\ifx\csname bibnamefont\endcsname\relax
  \def\bibnamefont#1{#1}\fi
\expandafter\ifx\csname bibfnamefont\endcsname\relax
  \def\bibfnamefont#1{#1}\fi
\expandafter\ifx\csname citenamefont\endcsname\relax
  \def\citenamefont#1{#1}\fi
\expandafter\ifx\csname url\endcsname\relax
  \def\url#1{\texttt{#1}}\fi
\expandafter\ifx\csname urlprefix\endcsname\relax\def\urlprefix{URL }\fi
\providecommand{\bibinfo}[2]{#2}
\providecommand{\eprint}[2][]{\url{#2}}

\bibitem[{\citenamefont{Feynman}(1982)}]{Feynman:1982}
\bibinfo{author}{\bibfnamefont{R.}~\bibnamefont{Feynman}},
  \bibinfo{journal}{Int. J. Theor. Phys.} \textbf{\bibinfo{volume}{21}},
  \bibinfo{pages}{467} (\bibinfo{year}{1982}).

\bibitem[{\citenamefont{Bloch et~al.}(2012)\citenamefont{Bloch, Dalibard, and
  Nascimbene}}]{Bloch:2012}
\bibinfo{author}{\bibfnamefont{I.}~\bibnamefont{Bloch}},
  \bibinfo{author}{\bibfnamefont{J.}~\bibnamefont{Dalibard}}, \bibnamefont{and}
  \bibinfo{author}{\bibfnamefont{S.}~\bibnamefont{Nascimbene}},
  \bibinfo{journal}{Nat. Phys.} \textbf{\bibinfo{volume}{8}},
  \bibinfo{pages}{267} (\bibinfo{year}{2012}).

\bibitem[{\citenamefont{Blatt and Roos}(2012)}]{Blatt:2012}
\bibinfo{author}{\bibfnamefont{R.}~\bibnamefont{Blatt}} \bibnamefont{and}
  \bibinfo{author}{\bibfnamefont{C.~F.} \bibnamefont{Roos}},
  \bibinfo{journal}{Nat. Phys.} \textbf{\bibinfo{volume}{8}},
  \bibinfo{pages}{277} (\bibinfo{year}{2012}).

\bibitem[{\citenamefont{Casanova et~al.}(2011)\citenamefont{Casanova, Lamata,
  Egusquiza, Gerritsma, Roos, Garc\'{\i}a-Ripoll, and Solano}}]{Casanova:2011b}
\bibinfo{author}{\bibfnamefont{J.}~\bibnamefont{Casanova}},
  \bibinfo{author}{\bibfnamefont{L.}~\bibnamefont{Lamata}},
  \bibinfo{author}{\bibfnamefont{I.~L.} \bibnamefont{Egusquiza}},
  \bibinfo{author}{\bibfnamefont{R.}~\bibnamefont{Gerritsma}},
  \bibinfo{author}{\bibfnamefont{C.~F.} \bibnamefont{Roos}},
  \bibinfo{author}{\bibfnamefont{J.~J.} \bibnamefont{Garc\'{\i}a-Ripoll}},
  \bibnamefont{and} \bibinfo{author}{\bibfnamefont{E.}~\bibnamefont{Solano}},
  \bibinfo{journal}{Phys. Rev. Lett.} \textbf{\bibinfo{volume}{107}},
  \bibinfo{pages}{260501} (\bibinfo{year}{2011}).

\bibitem[{\citenamefont{Garc\'{\i}a-\'Alvarez
  et~al.}(2015)\citenamefont{Garc\'{\i}a-\'Alvarez, Casanova, Mezzacapo,
  Egusquiza, Lamata, Romero, and Solano}}]{Garcia:2015}
\bibinfo{author}{\bibfnamefont{L.}~\bibnamefont{Garc\'{\i}a-\'Alvarez}},
  \bibinfo{author}{\bibfnamefont{J.}~\bibnamefont{Casanova}},
  \bibinfo{author}{\bibfnamefont{A.}~\bibnamefont{Mezzacapo}},
  \bibinfo{author}{\bibfnamefont{I.~L.} \bibnamefont{Egusquiza}},
  \bibinfo{author}{\bibfnamefont{L.}~\bibnamefont{Lamata}},
  \bibinfo{author}{\bibfnamefont{G.}~\bibnamefont{Romero}}, \bibnamefont{and}
  \bibinfo{author}{\bibfnamefont{E.}~\bibnamefont{Solano}},
  \bibinfo{journal}{Phys. Rev. Lett.} \textbf{\bibinfo{volume}{114}},
  \bibinfo{pages}{070502} (\bibinfo{year}{2015}).

\bibitem[{\citenamefont{Wilson}(1974)}]{Wilson:1974}
\bibinfo{author}{\bibfnamefont{K.~G.} \bibnamefont{Wilson}},
  \bibinfo{journal}{Phys. Rev. D} \textbf{\bibinfo{volume}{10}},
  \bibinfo{pages}{2445} (\bibinfo{year}{1974}).

\bibitem[{\citenamefont{Kogut}(1983)}]{Kogut:1983}
\bibinfo{author}{\bibfnamefont{J.~B.} \bibnamefont{Kogut}},
  \bibinfo{journal}{Rev. Mod. Phys.} \textbf{\bibinfo{volume}{55}},
  \bibinfo{pages}{775} (\bibinfo{year}{1983}).

\bibitem[{\citenamefont{Tagliacozzo et~al.}(2013)\citenamefont{Tagliacozzo,
  Celi, Zamora, and Lewenstein}}]{Tagliacozzo2013160}
\bibinfo{author}{\bibfnamefont{L.}~\bibnamefont{Tagliacozzo}},
  \bibinfo{author}{\bibfnamefont{A.}~\bibnamefont{Celi}},
  \bibinfo{author}{\bibfnamefont{A.}~\bibnamefont{Zamora}}, \bibnamefont{and}
  \bibinfo{author}{\bibfnamefont{M.}~\bibnamefont{Lewenstein}},
  \bibinfo{journal}{Annals of Physics} \textbf{\bibinfo{volume}{330}},
  \bibinfo{pages}{160 } (\bibinfo{year}{2013}).

\bibitem[{\citenamefont{Haegeman et~al.}(2010)\citenamefont{Haegeman, Cirac,
  Osborne, Verschelde, and Verstraete}}]{Haegeman:2010}
\bibinfo{author}{\bibfnamefont{J.}~\bibnamefont{Haegeman}},
  \bibinfo{author}{\bibfnamefont{J.~I.} \bibnamefont{Cirac}},
  \bibinfo{author}{\bibfnamefont{T.~J.} \bibnamefont{Osborne}},
  \bibinfo{author}{\bibfnamefont{H.}~\bibnamefont{Verschelde}},
  \bibnamefont{and}
  \bibinfo{author}{\bibfnamefont{F.}~\bibnamefont{Verstraete}},
  \bibinfo{journal}{Phys. Rev. Lett.} \textbf{\bibinfo{volume}{105}},
  \bibinfo{pages}{251601} (\bibinfo{year}{2010}).

\bibitem[{\citenamefont{Rico et~al.}(2014)\citenamefont{Rico, Pichler,
  Dalmonte, Zoller, and Montangero}}]{Rico:2014}
\bibinfo{author}{\bibfnamefont{E.}~\bibnamefont{Rico}},
  \bibinfo{author}{\bibfnamefont{T.}~\bibnamefont{Pichler}},
  \bibinfo{author}{\bibfnamefont{M.}~\bibnamefont{Dalmonte}},
  \bibinfo{author}{\bibfnamefont{P.}~\bibnamefont{Zoller}}, \bibnamefont{and}
  \bibinfo{author}{\bibfnamefont{S.}~\bibnamefont{Montangero}},
  \bibinfo{journal}{Phys. Rev. Lett.} \textbf{\bibinfo{volume}{112}},
  \bibinfo{pages}{201601} (\bibinfo{year}{2014}).

\bibitem[{\citenamefont{Tagliacozzo et~al.}(2014)\citenamefont{Tagliacozzo,
  Celi, and Lewenstein}}]{Tagliacozzo:2014}
\bibinfo{author}{\bibfnamefont{L.}~\bibnamefont{Tagliacozzo}},
  \bibinfo{author}{\bibfnamefont{A.}~\bibnamefont{Celi}}, \bibnamefont{and}
  \bibinfo{author}{\bibfnamefont{M.}~\bibnamefont{Lewenstein}},
  \bibinfo{journal}{Phys. Rev. X} \textbf{\bibinfo{volume}{4}},
  \bibinfo{pages}{041024} (\bibinfo{year}{2014}).

\bibitem[{\citenamefont{Ba\~nuls et~al.}(2015)\citenamefont{Ba\~nuls, Cichy,
  Cirac, Jansen, and Saito}}]{Banuls:2015}
\bibinfo{author}{\bibfnamefont{M.~C.} \bibnamefont{Ba\~nuls}},
  \bibinfo{author}{\bibfnamefont{K.}~\bibnamefont{Cichy}},
  \bibinfo{author}{\bibfnamefont{J.~I.} \bibnamefont{Cirac}},
  \bibinfo{author}{\bibfnamefont{K.}~\bibnamefont{Jansen}}, \bibnamefont{and}
  \bibinfo{author}{\bibfnamefont{H.}~\bibnamefont{Saito}},
  \bibinfo{journal}{Phys. Rev. D} \textbf{\bibinfo{volume}{92}},
  \bibinfo{pages}{034519} (\bibinfo{year}{2015}).

\bibitem[{\citenamefont{Pichler et~al.}(2016)\citenamefont{Pichler, Dalmonte,
  Rico, Zoller, and Montangero}}]{Pichler:2016}
\bibinfo{author}{\bibfnamefont{T.}~\bibnamefont{Pichler}},
  \bibinfo{author}{\bibfnamefont{M.}~\bibnamefont{Dalmonte}},
  \bibinfo{author}{\bibfnamefont{E.}~\bibnamefont{Rico}},
  \bibinfo{author}{\bibfnamefont{P.}~\bibnamefont{Zoller}}, \bibnamefont{and}
  \bibinfo{author}{\bibfnamefont{S.}~\bibnamefont{Montangero}},
  \bibinfo{journal}{Phys. Rev. X} \textbf{\bibinfo{volume}{6}},
  \bibinfo{pages}{011023} (\bibinfo{year}{2016}).

\bibitem[{\citenamefont{Martinez et~al.}(2016)\citenamefont{Martinez, Muschik,
  Schindler, Nigg, Erhard, Heyl, Hauke, Dalmonte, Monz, Zoller
  et~al.}}]{Martinez:2016}
\bibinfo{author}{\bibfnamefont{E.~A.} \bibnamefont{Martinez}},
  \bibinfo{author}{\bibfnamefont{C.~A.} \bibnamefont{Muschik}},
  \bibinfo{author}{\bibfnamefont{P.}~\bibnamefont{Schindler}},
  \bibinfo{author}{\bibfnamefont{D.}~\bibnamefont{Nigg}},
  \bibinfo{author}{\bibfnamefont{A.}~\bibnamefont{Erhard}},
  \bibinfo{author}{\bibfnamefont{M.}~\bibnamefont{Heyl}},
  \bibinfo{author}{\bibfnamefont{P.}~\bibnamefont{Hauke}},
  \bibinfo{author}{\bibfnamefont{M.}~\bibnamefont{Dalmonte}},
  \bibinfo{author}{\bibfnamefont{T.}~\bibnamefont{Monz}},
  \bibinfo{author}{\bibfnamefont{P.}~\bibnamefont{Zoller}},
  \bibnamefont{et~al.}, \bibinfo{journal}{Nature}
  \textbf{\bibinfo{volume}{534}}, \bibinfo{pages}{516} (\bibinfo{year}{2016}),
  \urlprefix\url{http://dx.doi.org/10.1038/nature18318}.

\bibitem[{\citenamefont{Kasamatsu et~al.}(2013)\citenamefont{Kasamatsu,
  Ichinose, and Matsui}}]{Higgs1}
\bibinfo{author}{\bibfnamefont{K.}~\bibnamefont{Kasamatsu}},
  \bibinfo{author}{\bibfnamefont{I.}~\bibnamefont{Ichinose}}, \bibnamefont{and}
  \bibinfo{author}{\bibfnamefont{T.}~\bibnamefont{Matsui}},
  \bibinfo{journal}{Phys. Rev. Lett.} \textbf{\bibinfo{volume}{111}},
  \bibinfo{pages}{115303} (\bibinfo{year}{2013}).

\bibitem[{\citenamefont{Edmonds et~al.}(2013)\citenamefont{Edmonds, Valiente,
  Juzeli\ifmmode~\bar{u}\else \={u}\fi{}nas, Santos, and
  \"Ohberg}}]{Edmonds:2013}
\bibinfo{author}{\bibfnamefont{M.~J.} \bibnamefont{Edmonds}},
  \bibinfo{author}{\bibfnamefont{M.}~\bibnamefont{Valiente}},
  \bibinfo{author}{\bibfnamefont{G.}~\bibnamefont{Juzeli\ifmmode~\bar{u}\else
  \={u}\fi{}nas}}, \bibinfo{author}{\bibfnamefont{L.}~\bibnamefont{Santos}},
  \bibnamefont{and} \bibinfo{author}{\bibfnamefont{P.}~\bibnamefont{\"Ohberg}},
  \bibinfo{journal}{Phys. Rev. Lett.} \textbf{\bibinfo{volume}{110}},
  \bibinfo{pages}{085301} (\bibinfo{year}{2013}).

\bibitem[{\citenamefont{Bazavov et~al.}(2015)\citenamefont{Bazavov, Meurice,
  Tsai, Unmuth-Yockey, and Zhang}}]{Higgs2}
\bibinfo{author}{\bibfnamefont{A.}~\bibnamefont{Bazavov}},
  \bibinfo{author}{\bibfnamefont{Y.}~\bibnamefont{Meurice}},
  \bibinfo{author}{\bibfnamefont{S.-W.} \bibnamefont{Tsai}},
  \bibinfo{author}{\bibfnamefont{J.}~\bibnamefont{Unmuth-Yockey}},
  \bibnamefont{and} \bibinfo{author}{\bibfnamefont{J.}~\bibnamefont{Zhang}},
  \bibinfo{journal}{Phys. Rev. D} \textbf{\bibinfo{volume}{92}},
  \bibinfo{pages}{076003} (\bibinfo{year}{2015}).

\bibitem[{\citenamefont{Gonz{\'a}lez-Cuadra
  et~al.}(2017)\citenamefont{Gonz{\'a}lez-Cuadra, Zohar, and Cirac}}]{Higgs3}
\bibinfo{author}{\bibfnamefont{D.}~\bibnamefont{Gonz{\'a}lez-Cuadra}},
  \bibinfo{author}{\bibfnamefont{E.}~\bibnamefont{Zohar}}, \bibnamefont{and}
  \bibinfo{author}{\bibfnamefont{J.~I.} \bibnamefont{Cirac}}
  (\bibinfo{year}{2017}), \eprint{arXiv:1702.05492}.

\bibitem[{\citenamefont{Horn}(1981)}]{HORN1981149}
\bibinfo{author}{\bibfnamefont{D.}~\bibnamefont{Horn}},
  \bibinfo{journal}{Physics Letters B} \textbf{\bibinfo{volume}{100}},
  \bibinfo{pages}{149 } (\bibinfo{year}{1981}).

\bibitem[{\citenamefont{Orland and Rohrlich}(1990)}]{ORLAND1990647}
\bibinfo{author}{\bibfnamefont{P.}~\bibnamefont{Orland}} \bibnamefont{and}
  \bibinfo{author}{\bibfnamefont{D.}~\bibnamefont{Rohrlich}},
  \bibinfo{journal}{Nuclear Physics B} \textbf{\bibinfo{volume}{338}},
  \bibinfo{pages}{647 } (\bibinfo{year}{1990}).

\bibitem[{\citenamefont{Chandrasekharan and
  Wiese}(1997)}]{CHANDRASEKHARAN1997455}
\bibinfo{author}{\bibfnamefont{S.}~\bibnamefont{Chandrasekharan}}
  \bibnamefont{and} \bibinfo{author}{\bibfnamefont{U.-J.} \bibnamefont{Wiese}},
  \bibinfo{journal}{Nuclear Physics B} \textbf{\bibinfo{volume}{492}},
  \bibinfo{pages}{455 } (\bibinfo{year}{1997}), ISSN \bibinfo{issn}{0550-3213}.

\bibitem[{\citenamefont{Kogut and Stephanov}(2003)}]{kogut2003phases}
\bibinfo{author}{\bibfnamefont{J.}~\bibnamefont{Kogut}} \bibnamefont{and}
  \bibinfo{author}{\bibfnamefont{M.}~\bibnamefont{Stephanov}},
  \emph{\bibinfo{title}{The Phases of Quantum Chromodynamics: From Confinement
  to Extreme Environments}}, Cambridge Monographs on Particle Physics, Nuclear
  Physics and Cosmology (\bibinfo{publisher}{Cambridge University Press},
  \bibinfo{year}{2003}).

\bibitem[{\citenamefont{Banerjee et~al.}(2012)\citenamefont{Banerjee, Dalmonte,
  M\"uller, Rico, Stebler, Wiese, and Zoller}}]{Banerjee:2012}
\bibinfo{author}{\bibfnamefont{D.}~\bibnamefont{Banerjee}},
  \bibinfo{author}{\bibfnamefont{M.}~\bibnamefont{Dalmonte}},
  \bibinfo{author}{\bibfnamefont{M.}~\bibnamefont{M\"uller}},
  \bibinfo{author}{\bibfnamefont{E.}~\bibnamefont{Rico}},
  \bibinfo{author}{\bibfnamefont{P.}~\bibnamefont{Stebler}},
  \bibinfo{author}{\bibfnamefont{U.-J.} \bibnamefont{Wiese}}, \bibnamefont{and}
  \bibinfo{author}{\bibfnamefont{P.}~\bibnamefont{Zoller}},
  \bibinfo{journal}{Phys. Rev. Lett.} \textbf{\bibinfo{volume}{109}},
  \bibinfo{pages}{175302} (\bibinfo{year}{2012}).

\bibitem[{\citenamefont{Marcos et~al.}(2013)\citenamefont{Marcos, Rabl, Rico,
  and Zoller}}]{Marcos:2013}
\bibinfo{author}{\bibfnamefont{D.}~\bibnamefont{Marcos}},
  \bibinfo{author}{\bibfnamefont{P.}~\bibnamefont{Rabl}},
  \bibinfo{author}{\bibfnamefont{E.}~\bibnamefont{Rico}}, \bibnamefont{and}
  \bibinfo{author}{\bibfnamefont{P.}~\bibnamefont{Zoller}},
  \bibinfo{journal}{Phys. Rev. Lett.} \textbf{\bibinfo{volume}{111}},
  \bibinfo{pages}{110504} (\bibinfo{year}{2013}).

\bibitem[{\citenamefont{Zohar et~al.}(2016)\citenamefont{Zohar, Cirac, and
  Reznik}}]{Zohar:2016}
\bibinfo{author}{\bibfnamefont{E.}~\bibnamefont{Zohar}},
  \bibinfo{author}{\bibfnamefont{J.~I.} \bibnamefont{Cirac}}, \bibnamefont{and}
  \bibinfo{author}{\bibfnamefont{B.}~\bibnamefont{Reznik}},
  \bibinfo{journal}{Rep. Prog. Phys.} \textbf{\bibinfo{volume}{79}},
  \bibinfo{pages}{014401} (\bibinfo{year}{2016}).

\bibitem[{\citenamefont{K\"uhn et~al.}(2014)\citenamefont{K\"uhn, Cirac, and
  Ba\~nuls}}]{Kuhn:2014}
\bibinfo{author}{\bibfnamefont{S.}~\bibnamefont{K\"uhn}},
  \bibinfo{author}{\bibfnamefont{J.~I.} \bibnamefont{Cirac}}, \bibnamefont{and}
  \bibinfo{author}{\bibfnamefont{M.-C.} \bibnamefont{Ba\~nuls}},
  \bibinfo{journal}{Phys. Rev. A} \textbf{\bibinfo{volume}{90}},
  \bibinfo{pages}{042305} (\bibinfo{year}{2014}).

\bibitem[{\citenamefont{Buyens et~al.}(2017)\citenamefont{Buyens, Montangero,
  Haegeman, Verstraete, and Van~Acoleyen}}]{Buyens:2017}
\bibinfo{author}{\bibfnamefont{B.}~\bibnamefont{Buyens}},
  \bibinfo{author}{\bibfnamefont{S.}~\bibnamefont{Montangero}},
  \bibinfo{author}{\bibfnamefont{J.}~\bibnamefont{Haegeman}},
  \bibinfo{author}{\bibfnamefont{F.}~\bibnamefont{Verstraete}},
  \bibnamefont{and}
  \bibinfo{author}{\bibfnamefont{K.}~\bibnamefont{Van~Acoleyen}},
  \bibinfo{journal}{arXiv:1702.08838v1}  (\bibinfo{year}{2017}).

\bibitem[{\citenamefont{Kasper et~al.}(2017)\citenamefont{Kasper, Hebenstreit,
  Jendrzejewski, Oberthaler, and Berges}}]{Kasper:2017}
\bibinfo{author}{\bibfnamefont{V.}~\bibnamefont{Kasper}},
  \bibinfo{author}{\bibfnamefont{F.}~\bibnamefont{Hebenstreit}},
  \bibinfo{author}{\bibfnamefont{F.}~\bibnamefont{Jendrzejewski}},
  \bibinfo{author}{\bibfnamefont{M.~K.} \bibnamefont{Oberthaler}},
  \bibnamefont{and} \bibinfo{author}{\bibfnamefont{J.}~\bibnamefont{Berges}},
  \bibinfo{journal}{New Journal of Physics} \textbf{\bibinfo{volume}{19}},
  \bibinfo{pages}{023030} (\bibinfo{year}{2017}).

\bibitem[{\citenamefont{Fradkin}(2013)}]{fradkin2013field}
\bibinfo{author}{\bibfnamefont{E.}~\bibnamefont{Fradkin}},
  \emph{\bibinfo{title}{Field Theories of Condensed Matter Physics}}, Field
  Theories of Condensed Matter Physics (\bibinfo{publisher}{Cambridge
  University Press}, \bibinfo{year}{2013}).

\bibitem[{\citenamefont{Micheli et~al.}(2004)\citenamefont{Micheli, Daley,
  Jaksch, and Zoller}}]{Micheli:2004}
\bibinfo{author}{\bibfnamefont{A.}~\bibnamefont{Micheli}},
  \bibinfo{author}{\bibfnamefont{A.~J.} \bibnamefont{Daley}},
  \bibinfo{author}{\bibfnamefont{D.}~\bibnamefont{Jaksch}}, \bibnamefont{and}
  \bibinfo{author}{\bibfnamefont{P.}~\bibnamefont{Zoller}},
  \bibinfo{journal}{Phys. Rev. Lett.} \textbf{\bibinfo{volume}{93}},
  \bibinfo{pages}{140408} (\bibinfo{year}{2004}).

\bibitem[{\citenamefont{Gerritsma et~al.}(2012)\citenamefont{Gerritsma,
  Negretti, Doerk, Idziaszek, Calarco, and Schmidt-Kaler}}]{Gerritsma:2012}
\bibinfo{author}{\bibfnamefont{R.}~\bibnamefont{Gerritsma}},
  \bibinfo{author}{\bibfnamefont{A.}~\bibnamefont{Negretti}},
  \bibinfo{author}{\bibfnamefont{H.}~\bibnamefont{Doerk}},
  \bibinfo{author}{\bibfnamefont{Z.}~\bibnamefont{Idziaszek}},
  \bibinfo{author}{\bibfnamefont{T.}~\bibnamefont{Calarco}}, \bibnamefont{and}
  \bibinfo{author}{\bibfnamefont{F.}~\bibnamefont{Schmidt-Kaler}},
  \bibinfo{journal}{Phys. Rev. Lett.} \textbf{\bibinfo{volume}{109}},
  \bibinfo{pages}{080402} (\bibinfo{year}{2012}).

\bibitem[{\citenamefont{Schurer et~al.}(2016)\citenamefont{Schurer, Gerritsma,
  Schmelcher, and Negretti}}]{Schurer:2016}
\bibinfo{author}{\bibfnamefont{J.~M.} \bibnamefont{Schurer}},
  \bibinfo{author}{\bibfnamefont{R.}~\bibnamefont{Gerritsma}},
  \bibinfo{author}{\bibfnamefont{P.}~\bibnamefont{Schmelcher}},
  \bibnamefont{and} \bibinfo{author}{\bibfnamefont{A.}~\bibnamefont{Negretti}},
  \bibinfo{journal}{Phys. Rev. A} \textbf{\bibinfo{volume}{93}},
  \bibinfo{pages}{063602} (\bibinfo{year}{2016}).

\bibitem[{\citenamefont{Negretti et~al.}(2014)\citenamefont{Negretti,
  Gerritsma, Idziaszek, Schmidt-Kaler, and Calarco}}]{Negretti:2014}
\bibinfo{author}{\bibfnamefont{A.}~\bibnamefont{Negretti}},
  \bibinfo{author}{\bibfnamefont{R.}~\bibnamefont{Gerritsma}},
  \bibinfo{author}{\bibfnamefont{Z.}~\bibnamefont{Idziaszek}},
  \bibinfo{author}{\bibfnamefont{F.}~\bibnamefont{Schmidt-Kaler}},
  \bibnamefont{and} \bibinfo{author}{\bibfnamefont{T.}~\bibnamefont{Calarco}},
  \bibinfo{journal}{Phys. Rev. B} \textbf{\bibinfo{volume}{90}},
  \bibinfo{pages}{155426} (\bibinfo{year}{2014}).

\bibitem[{\citenamefont{Girardeau and Olshanii}(2004)}]{Girardeau:2004}
\bibinfo{author}{\bibfnamefont{M.~D.} \bibnamefont{Girardeau}}
  \bibnamefont{and} \bibinfo{author}{\bibfnamefont{M.}~\bibnamefont{Olshanii}},
  \bibinfo{journal}{Phys. Rev. A} \textbf{\bibinfo{volume}{70}},
  \bibinfo{pages}{023608} (\bibinfo{year}{2004}).

\bibitem[{\citenamefont{{Valiente} and {Zinner}}(2015)}]{Valiente:2015}
\bibinfo{author}{\bibfnamefont{M.}~\bibnamefont{{Valiente}}} \bibnamefont{and}
  \bibinfo{author}{\bibfnamefont{N.~T.} \bibnamefont{{Zinner}}},
  \bibinfo{journal}{Few-Body Systems} \textbf{\bibinfo{volume}{56}},
  \bibinfo{pages}{845} (\bibinfo{year}{2015}).

\bibitem[{\citenamefont{Joger et~al.}(2014)\citenamefont{Joger, Negretti, and
  Gerritsma}}]{Joger:2014}
\bibinfo{author}{\bibfnamefont{J.}~\bibnamefont{Joger}},
  \bibinfo{author}{\bibfnamefont{A.}~\bibnamefont{Negretti}}, \bibnamefont{and}
  \bibinfo{author}{\bibfnamefont{R.}~\bibnamefont{Gerritsma}},
  \bibinfo{journal}{Phys. Rev. A} \textbf{\bibinfo{volume}{89}},
  \bibinfo{pages}{063621} (\bibinfo{year}{2014}).

\bibitem[{\citenamefont{Leibfried et~al.}(2003)\citenamefont{Leibfried, Blatt,
  Monroe, and Wineland}}]{Leibfried:2003}
\bibinfo{author}{\bibfnamefont{D.}~\bibnamefont{Leibfried}},
  \bibinfo{author}{\bibfnamefont{R.}~\bibnamefont{Blatt}},
  \bibinfo{author}{\bibfnamefont{C.}~\bibnamefont{Monroe}}, \bibnamefont{and}
  \bibinfo{author}{\bibfnamefont{D.}~\bibnamefont{Wineland}},
  \bibinfo{journal}{Rev. Mod. Phys.} \textbf{\bibinfo{volume}{75}},
  \bibinfo{pages}{281} (\bibinfo{year}{2003}).

\bibitem[{\citenamefont{Cook et~al.}(1985)\citenamefont{Cook, Shankland, and
  Wells}}]{Cook:1985}
\bibinfo{author}{\bibfnamefont{R.~J.} \bibnamefont{Cook}},
  \bibinfo{author}{\bibfnamefont{D.~G.} \bibnamefont{Shankland}},
  \bibnamefont{and} \bibinfo{author}{\bibfnamefont{A.~L.} \bibnamefont{Wells}},
  \bibinfo{journal}{Phys. Rev. A} \textbf{\bibinfo{volume}{31}},
  \bibinfo{pages}{564} (\bibinfo{year}{1985}).

\bibitem[{\citenamefont{Cetina et~al.}(2012)\citenamefont{Cetina, Grier, and
  Vuleti\ifmmode~\acute{c}\else \'{c}\fi{}}}]{Cetina:2012}
\bibinfo{author}{\bibfnamefont{M.}~\bibnamefont{Cetina}},
  \bibinfo{author}{\bibfnamefont{A.~T.} \bibnamefont{Grier}}, \bibnamefont{and}
  \bibinfo{author}{\bibfnamefont{V.}~\bibnamefont{Vuleti\ifmmode~\acute{c}\else
  \'{c}\fi{}}}, \bibinfo{journal}{Phys. Rev. Lett.}
  \textbf{\bibinfo{volume}{109}}, \bibinfo{pages}{253201}
  (\bibinfo{year}{2012}).

\bibitem[{\citenamefont{Tscherbul et~al.}(2016)\citenamefont{Tscherbul, Brumer,
  and Buchachenko}}]{Tscherbul:2016}
\bibinfo{author}{\bibfnamefont{T.~V.} \bibnamefont{Tscherbul}},
  \bibinfo{author}{\bibfnamefont{P.}~\bibnamefont{Brumer}}, \bibnamefont{and}
  \bibinfo{author}{\bibfnamefont{A.~A.} \bibnamefont{Buchachenko}},
  \bibinfo{journal}{Phys. Rev. Lett.} \textbf{\bibinfo{volume}{117}},
  \bibinfo{pages}{143201} (\bibinfo{year}{2016}).

\bibitem[{\citenamefont{Joger et~al.}(2017)\citenamefont{Joger, F\"urst, Ewald,
  Feldker, Tomza, and Gerritsma}}]{Joger:2017}
\bibinfo{author}{\bibfnamefont{J.}~\bibnamefont{Joger}},
  \bibinfo{author}{\bibfnamefont{H.}~\bibnamefont{F\"urst}},
  \bibinfo{author}{\bibfnamefont{N.}~\bibnamefont{Ewald}},
  \bibinfo{author}{\bibfnamefont{T.}~\bibnamefont{Feldker}},
  \bibinfo{author}{\bibfnamefont{M.}~\bibnamefont{Tomza}}, \bibnamefont{and}
  \bibinfo{author}{\bibfnamefont{R.}~\bibnamefont{Gerritsma}},
  \bibinfo{journal}{arXiv:1707.01729}  (\bibinfo{year}{2017}).

\bibitem[{\citenamefont{Abramowitz and Stegun}(1992)}]{Abramowitz:1992}
\bibinfo{editor}{\bibfnamefont{M.}~\bibnamefont{Abramowitz}} \bibnamefont{and}
  \bibinfo{editor}{\bibfnamefont{I.~A.} \bibnamefont{Stegun}}, eds.,
  \emph{\bibinfo{title}{Handbook of mathematical functions with formulas,
  graphs, and mathematical tables}} (\bibinfo{publisher}{Dover Publications,
  Inc., New York}, \bibinfo{year}{1992}), \bibinfo{note}{reprint of the 1972
  edition}.

\bibitem[{\citenamefont{Busch et~al.}(1998)\citenamefont{Busch, Englert,
  Rza{\.{z}}ewski, and Wilkens}}]{Busch:1998}
\bibinfo{author}{\bibfnamefont{T.}~\bibnamefont{Busch}},
  \bibinfo{author}{\bibfnamefont{B.-G.} \bibnamefont{Englert}},
  \bibinfo{author}{\bibfnamefont{K.}~\bibnamefont{Rza{\.{z}}ewski}},
  \bibnamefont{and} \bibinfo{author}{\bibfnamefont{M.}~\bibnamefont{Wilkens}},
  \bibinfo{journal}{Foundations of Physics} \textbf{\bibinfo{volume}{28}},
  \bibinfo{pages}{549} (\bibinfo{year}{1998}).

\bibitem[{\citenamefont{Brouzos et~al.}(2015)\citenamefont{Brouzos, Streltsov,
  Negretti, Said, Caneva, Montangero, and Calarco}}]{Brouzos:2015}
\bibinfo{author}{\bibfnamefont{I.}~\bibnamefont{Brouzos}},
  \bibinfo{author}{\bibfnamefont{A.~I.} \bibnamefont{Streltsov}},
  \bibinfo{author}{\bibfnamefont{A.}~\bibnamefont{Negretti}},
  \bibinfo{author}{\bibfnamefont{R.~S.} \bibnamefont{Said}},
  \bibinfo{author}{\bibfnamefont{T.}~\bibnamefont{Caneva}},
  \bibinfo{author}{\bibfnamefont{S.}~\bibnamefont{Montangero}},
  \bibnamefont{and} \bibinfo{author}{\bibfnamefont{T.}~\bibnamefont{Calarco}},
  \bibinfo{journal}{Phys. Rev. A} \textbf{\bibinfo{volume}{92}},
  \bibinfo{pages}{062110} (\bibinfo{year}{2015}).

\end{thebibliography}
\end{document}